%% file: 3_Main.tex
\documentclass[sn-mathphys,iicol,Numbered]{sn-jnl}
\usepackage{graphicx}%
\usepackage{multirow}%
\usepackage{amsmath,amssymb,amsfonts}%
\usepackage{amsthm}%
\usepackage{mathrsfs}%
\usepackage[title]{appendix}%
\usepackage{xcolor}%
\usepackage{textcomp}%
\usepackage{manyfoot}%
\usepackage{booktabs}%
\usepackage{algorithm}%
\usepackage{algorithmicx}%
\usepackage{algpseudocode}%
\usepackage{listings}%
\usepackage{chemformula}
\usepackage{geometry} 

\unnumbered 

\begin{document}

\title{Electron hopping induced phonon pumping in opto-mechanical molecular nanocavities}

\author[1]{\fnm{Yu} \sur{Bai}}
\equalcont{These authors contributed equally to this work.}

\author[1]{\fnm{Ilya} \sur{Razdolski}}
\equalcont{These authors contributed equally to this work.}

\author[2]{\fnm{Zhizi} \sur{Guan}}

\author[3]{\fnm{Ping} \sur{Tang}}

\author[1]{\fnm{Xiu} \sur{Liang}}

\author[2]{\fnm{David J.} \sur{Srolovitz}}

\author[4]{\fnm{Anatoly V.} \sur{Zayats}} 

\author*[1]{\fnm{Dangyuan} \sur{Lei}}\email{dangylei@cityu.edu.hk}

\affil[1]{\orgdiv{Department of Materials Science and Engineering, Department of Physics, Center for Functional Photonics, Hong Kong Branch of National Precious Metals Material Engineering Research Centre, and Hong Kong Institute of Clean Energy}, \orgname{City University of Hong Kong}, \orgaddress{\street{83 Tat Chee Avenue}, \city{Hong Kong S.A.R.}, \postcode{999077}, \country{China}}}

\affil[2]{\orgdiv{Department of Mechanical Engineering}, \orgname{The University of Hong Kong}, \orgaddress{\street{Pokfulam Road}, \city{Hong Kong S.A.R.}, \postcode{999077},  \country{China}}}

\affil[3]{\orgdiv{School of Physics and Optoelectronic Engineering, Key Laboratory of Photonics Technology for Integrated Sensing and Communication of Ministry of Education}, \orgname{Guangdong University of Technology}, \orgaddress{\city{Guangzhou}, \postcode{510006}, \country{China}}}

\affil[4]{\orgdiv{Department of Physics and London Centre for Nanotechnology}, \orgname{King\'s College London}, \orgaddress{\street{Strand}, \city{London}, \postcode{WC2R 2LS}, \country{UK}}}


\abstract{
Plasmonic molecular nanojunctions exhibit opto-mechanical coupling at the nanoscale, enabling intertwined optical, vibrational and electronic phenomena. 
Here, we demonstrate plasmon-mediated phonon pumping, driven by inelastic electron hopping in conductive molecules, which results in strong Raman nonlinearity at the light intensities almost three orders of magnitude lower than in the conventional opto-mechanical systems and up to four-fold enhancement of the effective Raman polarizability due to vibrational electron-phonon coupling, as confirmed by the significant increase in anti-Stokes Raman scattering intensity, indicating enhanced vibrational occupancy. We also developed a microscopic framework of opto-mechanical electron-phonon coupling in molecular nanojunctions based on the Marcus electron hopping. Systematically varying electrical conductance of the molecules in the junction and laser intensity, we observed the transition between a photo-assisted tunneling regime and an electron hopping process. Our findings provide a microscopic description for vibrational, optical, and electronic phenomena in plasmonic nanocavities important for efficient phonon lasing, representing the first attempt to exploit conductive molecules as
quantum-mechanical oscillators.
}

\keywords{inelastic electron scattering, Marcus current, molecular nanojunctions, Raman spectroscopy, plasmonics}



\maketitle
Integration of functional molecules into nanophotonic systems is a promising and intensely studied concept in modern nano- and opto-electronics \cite{BatraNanoLett2013, CapozziNatNano2015,CanevaNatNano2018, LiNatNano2018,ChenScience2021,XomalisScience2021,wang2018reactive}. These systems offer unique opportunities to leverage molecular functionalities while exploiting nanoscale photonic platforms for manipulating light-matter interactions. Demonstrating uniquely strong light confinement, plasmonic molecular nanocavities with ultrasmall nanogaps ($\lesssim 1$~nm) constitute an ideal platform for studying novel opto-mechanical coupling effects 
\cite{ThossJCP2018,ZhuNatComm2014,RoelliNatNano2016} through surface-enhanced Raman scattering (SERS), tip-enhanced Raman scattering and scanning tunneling microscopy, enabling unprecedented sensitivity for probing molecular vibrations~\cite{SchmidtACSNano2016,cirera2022charge}. Mapping the long-known formalism for the parametric driving of optical microcavities \cite{KippenbergPRL2005,SchliesserPRL2006,TeufelPRL2008} onto plasmonic molecular junctions, dynamical back-action, parametric instability, phonon amplification and cooling, up-conversion and optical spring effects were predicted and observed \cite{SchmidtACSNano2016, BenzScience2016, LombardiPRX2018, ChenScience2021, JakobNatComm2023,xu2022phononic,hu2019closely}. Moreover, opto-mechanical coupling has been identified as a promising route to achieve phonon lasing through radiation pressure \cite{kippenberg2008cavity,vahala2009phonon,czerniuk2014lasing}. Importantly, this amplification can be further enhanced by exploiting electron-phonon coupling in molecular nanojunction systems \cite{ness2001coherent,seldenthuis2008vibrational}.

Yet, understanding of the underlying physical mechanisms for the opto-mechanical coupling requires bridging the gap between the optical and electronic pictures. Whereas radiation pressure is arguably responsible for the coupling in microcavities \cite{AspelmeyerRMP2014}, its interplay with the electron tunneling current in molecular plasmonic nanojunctions and whether the same mechanism dominates, have not yet been clearly identified. Inelastic electron spectroscopy is typically employed to probe vibrational resonances of the molecules \cite{ReedMaterials2014, EickhoffPRB2020} while vibrational instabilities are attributed to the laser-induced phonon pumping \cite{KneippPRL1996, SchmidtACSNano2016, LombardiPRX2018}. The intimate relation between inelastic electron tunneling and nano-optics \cite{wang2018reactive, qian2018efficient, MuniainPRX2024} is further emphasized in the non-thermal electron- or photo-assisted tunneling (PAT) effects \cite{ThonAPA2004, StolzNanoLett2014, FungNanoLett2017,kos2021quantum}.

In this work, we demonstrate that inelastic electron hopping in conductive molecules within plasmonic nanocavities provides an effective approach to photo-induced charge transport across molecular nanojunctions, resulting in phonon excitation and the related enhancement of Raman scattering. We also developed an overarching microscopic model for the opto-mechanical coupling phenomena in plasmonic nanogaps. Emphasizing the key role of inelastic electron hopping, vibrational mode-specific electron-phonon coupling and molecular conductance, this work provides new insights into nanoscale opto-mechanical coupling mechanisms and lays the theoretical groundwork for plasmon-induced phonon lasing, highlighting its potential implications for both quantum plasmonics and next-generation molecular devices.

\section*{Results}

\begin{centering}
\begin{figure*}[ht]%
\centering
\includegraphics[width=0.95\textwidth]{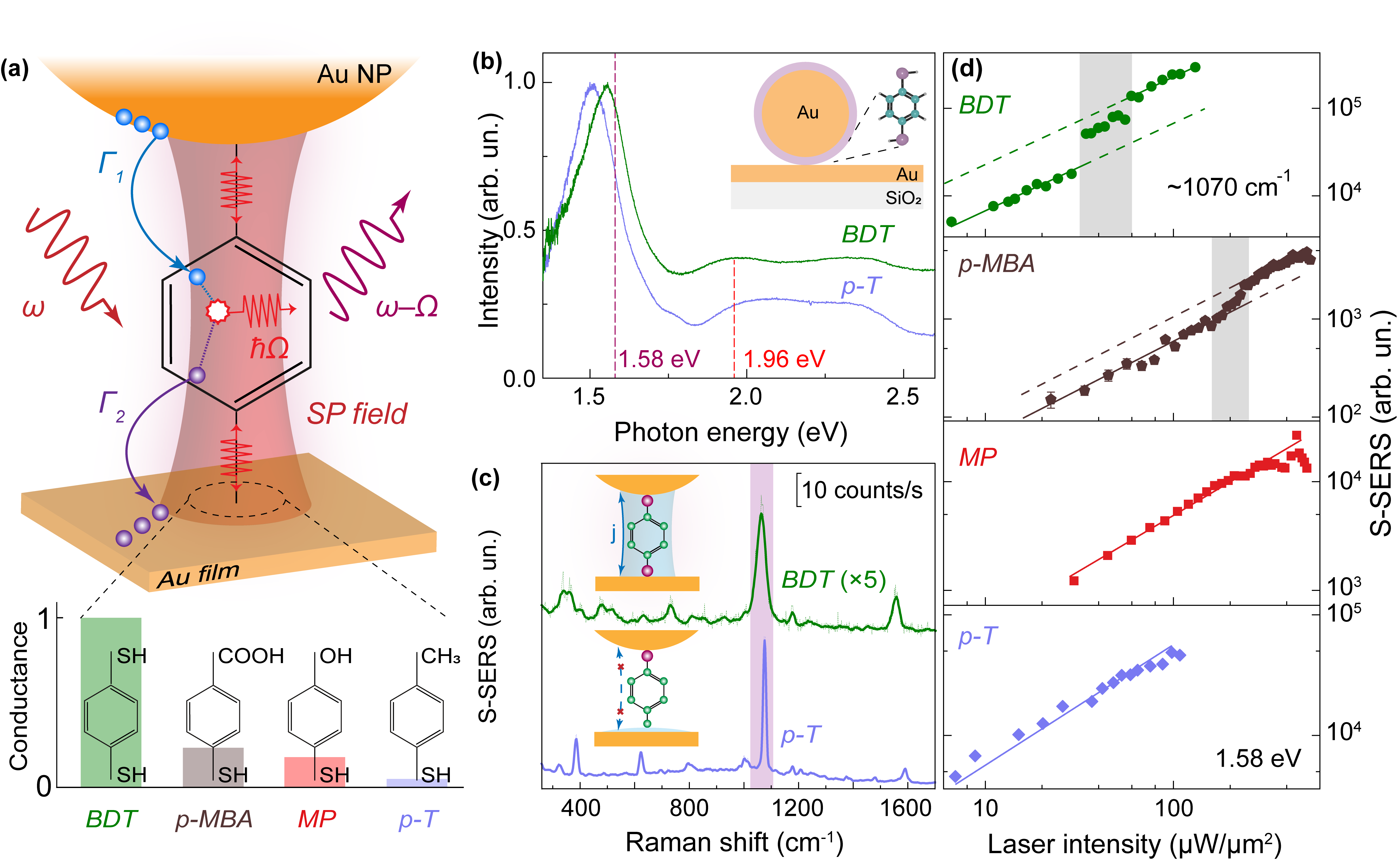}
\caption{
{\bf Raman spectroscopy of inelastic electron hopping in molecular plasmonic nanojunctions.} 
(a) Schematic of a Raman process in a molecule in a plasmonic junction. Illuminating light excites a surface plasmon in the NPoM (SP field, red shaded area). The Stokes photons emitted through Raman scattering at $\omega\rightarrow\omega-\Omega$ bear the imprint of the inelastic electron dynamics. The bottom inset compares the relative $dc$ conductance of the four studied molecules (see also  Supplementary Material, Fig.~S10) and shows their atomic structure.
(b) Single-particle dark-field scattering spectra of the NPoM plasmonic junctions with BDT and p-T molecules with the highest and the lowest conductance studied. The vertical dashed lines indicate the photon energies used for resonant (1.58~eV) and non-resonant (1.96~eV) Raman excitation. The inset shows the geometry of the molecular NPoM nanojunctions. 
(c) Stokes Raman spectra of the plasmonic nanojunctions with BDT and p-T molecules under on-resonance (1.58~eV)  excitation. The spectra are offset vertically for clarity. The insets show schematics of the respective molecules and their bonding to Au: symmetric bonding of BDT enables electric current across the molecular nanojunction (blue shaded) whereas in p-T, the CH$_3$ coupling to Au is inefficient and thus the current is suppressed. The purple shaded area indicates the vibrational mode at $\hbar\Omega\sim1070$~cm$^{-1}$ which is analyzed in (d). The 1070 cm$^{-1}$ vibrational mode of BDT exhibits a broader FWHM compared to p-T, consistent with the previous studies on dithiol molecules in plasmonic systems \cite{cui2018molecular,suzuki2016effect,caligiuri2020biodegradable}. The FWHM of BDT, typically 10–20 cm$^{-1}$, yet remains constant across laser intensities (Fig. S7), indicating an intrinsic molecular property rather than a laser-power-dependent effect.
(d) Excitation intensity dependences of the integrated S-SERS response at $\hbar\Omega\sim1070$~cm$^{-1}$ measured with the NPoMs for four studied molecules. Only nanojunctions with conductive BDT and p-MBA molecules exhibit nonlinear Raman behavior. The solid and dashed lines indicate linear dependence. The gray shaded areas highlight the nonlinear regime in the BDT and p-MBA nanojunctions response.
}
\label{fig:1}
\end{figure*}
\end{centering}

\subsection*{Conductive molecular plasmonic junctions}\label{sec12}

Opto-mechanical coupling is studied in plasmonic molecular nanojunctions, based on a nanoparticle-on-mirror (NPoM) geometry, which provides strong field enhancement in the nanogap region \cite{AravindJPC1982, BaumbergNatMater2019, HuNanop2022}. 80 nm-diameter Au nanospheres were functionalized by coating with a self-assembled molecular monolayer (see Methods and Figs.~S1--S3 for the details of the fabrication) and deposited onto a smooth Au film, thus forming ultrasmall (0.7~nm separation) plasmonic nanocavities (Fig. \ref{fig:1}a). To tune the electrical conductance in the nanojunction, four aromatic thiol molecules with similar optical but distinct electronic properties were used: 1,4-benzenedithiol (BDT), 4-mercaptobenzoic acid (p-MBA), 4-mercaptophenol (MP), and p-Thiocresol (p-T). The thiol group ensures strong chemical bonding to gold \cite{thomas2018acid}. Dominating the electron transport in aromatic bridges, this bonding provides a conductive link through an overlap of the $\pi$-orbitals of the phenyl ring with the wavefunctions of electrons in gold \cite{cui2018molecular,benz2015nanooptics}. However, while in BDT molecules a \ch{Au-S} covalent bond formed at both sides enables efficient electron transport across the nanojunction, in the other molecules (p-MBA, MP, and p-T), $\ch{-OH}$ and $\ch{-CH}_{3}$ groups cannot form robust bonds to gold \cite{cui2018molecular,park2010charge}. This key difference is responsible for the variations of the $dc$ conductance and optical extinction of the nanojunctions, whose localized surface plasmon (LSP) resonance is observed at around 1.5~eV (Fig. \ref{fig:1}b). This can be understood within a circuit model: molecules with larger conductance decrease the $R-C$ time constant, which blueshifts and broadens the NPoM LSP~\cite{benz2015generalized,cui2018molecular}.

\subsection*{Nonlinear Raman scattering}

\begin{figure}[ht]%
\centering
\includegraphics[width=0.45\textwidth]{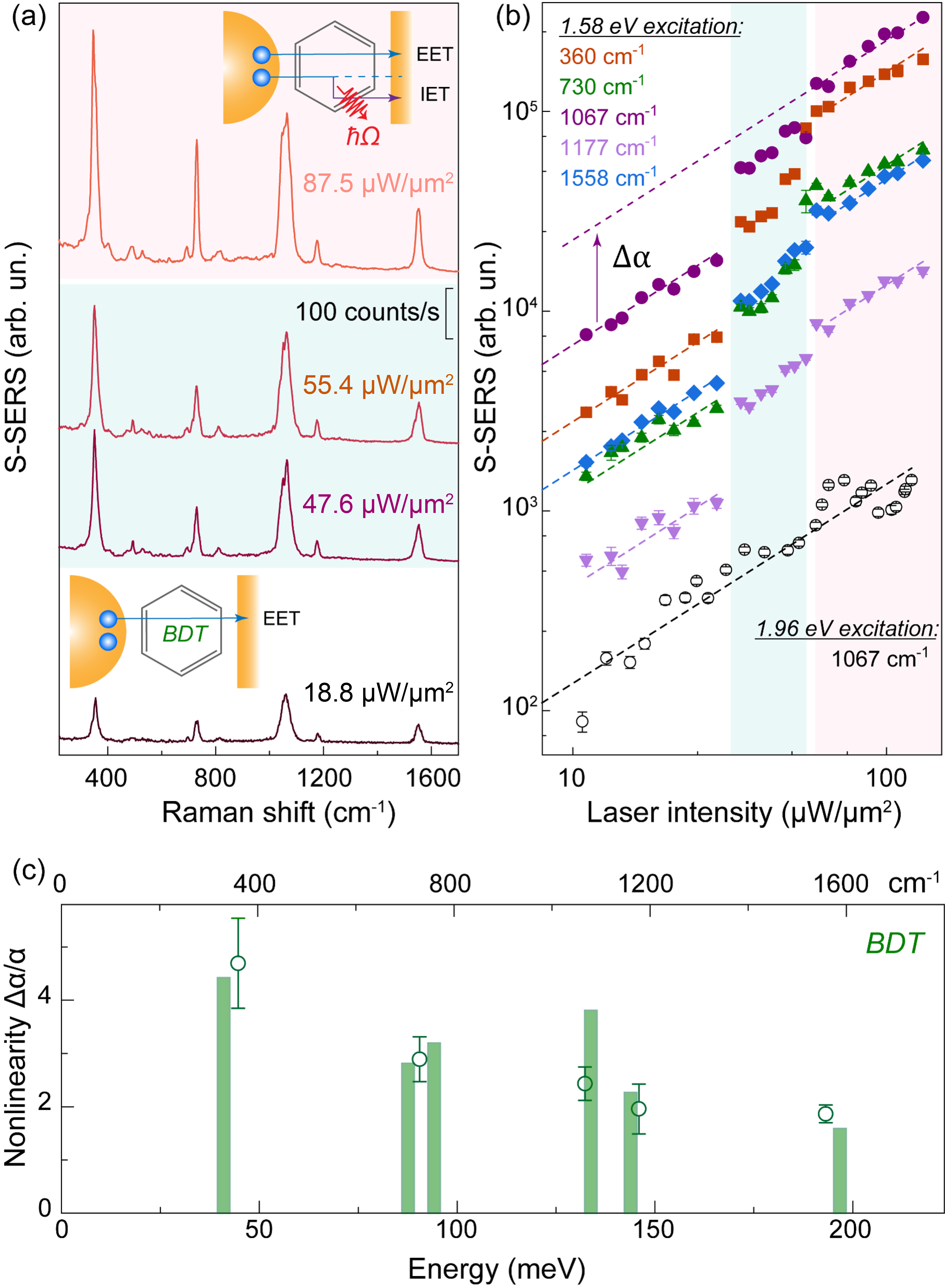}
\caption{
{\bf Nonlinear Raman scattering regime in conductive NPoM nanojunctions with BDT molecules.} 
(a) S-SERS spectra of BDT molecules in NPoM nanojunctions in the low-intensity (white), nonlinear (teal) and high-intensity (magenta) regimes. The laser intensity in $\mu W/\mu m^2$ indicated at each spectrum. The spectra are offset vertically for clarity. The insets schematically illustrate the activated electron transfer processes: in the low-intensity regime, the tunneling is elastic (EET) whereas in the high-intensity one, inelastic electron tunneling (IET) emerges, resonantly pumping phonons with energies $\hbar\Omega$.
(b) Excitation intensity dependences of the integrated (excluding the background) S-SERS intensity of the BDT vibrational modes under on-resonance (1.58~eV, colored full dots) and off-resonance (1.96~eV, black open dots) excitation. 
The dashed lines indicate linear dependence between the laser intensity and the S-SERS intensity.
The shaded areas highlight different excitation intensity regimes as in (a).
(c) Opto-mechanical relative enhancement factors of the Raman polarizability $\Delta\alpha/\alpha$ for different vibrational modes, extracted from bi-linear fitting of the data in panel (b). Bars represent the theoretical electron-phonon coupling for the vibrational modes in BDT molecules obtained in Ref.~\cite{seldenthuisACSNano2008}. 
}
\label{fig:2}
\end{figure}

In striking contrast, the plasmonic nanojunctions with either p-T or MP molecules which have low conductivity do not demonstrate nonlinear behaviour with a linear trend under both resonant  (1.58~eV) and non-resonant (1.96~eV) excitations (Figs. S9 and S10). For BDT and p-MBA plasmonic nanojunctions, we did not observe significant modification of the Raman spectra in the nonlinear region (Fig. \ref{fig:2}a), in contrast to the predictions of the widely used model for opto-mechanical coupling in plasmonic cavities \cite{RoelliNatNano2016}. First, the linewidths of the observed Raman modes exhibit no reduction in the nonlinear regime (Fig. S7), unlike in the conventional opto-mechanical coupling model, which expects the coupling-driven amplification factor to negate the intrinsic phonon damping when the parametric vibrational instability is reached~\cite{LombardiPRX2018}. Second, the nonlinear regime is reached at significantly (2--3 orders of magnitude) lower excitation laser intensity levels than reported in other works \cite{SchmidtACSNano2016, LombardiPRX2018}.
Moreover, because the coupling strength increases with the plasmonic near-field enhancement, the conventional opto-mechanical model would expect a lower nonlinearity threshold in the p-MBA NPoM system with smaller conductivity than BDT. Even lower excitation intensities would be expected to reach nonlinearity in lower-conductive MP and p-T nanojunctions, opposite to our experimental findings. 

Lastly, and most importantly, whereas the Raman response to the off-resonant excitation (1.96~eV) remains linear until the molecular damage threshold (Fig. S10). In contrast, at the LSP excitation (1.58~eV), all vibrational modes of BDT and p-MBA molecules show nonlinear behavior in the same intensity range. These observations are inconsistent with the conventional opto-mechanical coupling model, where the threshold for the instability is tied to the coupling strength which depends on a particular vibrational mode. 

At high excitation intensities, the electric field gradient at the molecule-metal contact reportedly pulls the atoms at the metal surface from their original positions \cite{carnegie2018room,lin2022optical}. In such picocavity geometry, the electric field enhancement can be further increased \cite{BenzScience2016}, potentially boosting the Raman response. However, time-dependent measurements of S-SERS spectra of the BDT nanojunctions in the nonlinear regime ($50~\mu$W/$\mu$m$^{2}$) did not show the instabilities related to atom movements (Fig. S8).  The S-SERS spectra of BDT maintain consistent peak positions and linewidths across the entire laser power range (Fig. S7). These findings rule out conformational changes or alterations in the metal structure as the origin of the observed nonlinear effects.

On the other hand, the S-SERS signal $R_S$ at elevated laser intensities $I$ can be described by variations of the Raman polarizability $\alpha$: $R_S\propto\alpha I$.  If stimulated phonon emission is present (indicative of the S-shaped intensity dependence) due to inelastic electron scattering (IES, see insets in Fig.~2a), vibrational mode-specific nonlinearity $\Delta\alpha(\Omega)$ will emerge. Strong nonlinearity above the threshold with $\Delta\alpha/\alpha\sim2-4$ (Fig.~2c) provides stronger amplification of the Raman response than the 15-35\% enhancement in the low-conductance counterparts \cite{XomalisScience2021}.
Moreover, an observed correlation between the experimental results and the calculated electron-phonon coupling $g_{e-ph}(\Omega)$ in BDT molecules~\cite{seldenthuisACSNano2008} further corroborates the key contribution of the inelastic electron hopping in the plasmonic nanojunctions to opto-mechanical coupling.

\subsection{Photo-assisted molecular conductance}

As an electronic property of the nanojunctions, molecular conductivity can be incorporated into an all-round model for optical and electronic properties of a quantum molecular oscillator through the PAT framework. 
We begin with the calculations of the electron transport across the molecular nanojunctions. Employing the non-equilibrium Green’s function - density functional theory (NEGF-DFT) method implemented in tranSIESTA \cite{soler2002siesta,brandbyge2002density}, we determine the electronic transmittance function $T(\mathcal{E})$, where $\mathcal{E}$ is the electron energy. Our calculations may be directly compared with the experimental molecule-Au(111) junctions by first performing geometry optimizations via DFT in the Perdew–Burke–Ernzerhof generalized gradient approximation  \cite{kresse1996efficient,perdew1996generalized,blochl1994projector} (see Supplementary Material-DFT calculations). The relaxed atomic structures are displayed in Fig.~S8 together with the distance between the end of the molecule and the Au plane.

\begin{figure}[t]%
\centering
\includegraphics[width=0.47\textwidth]{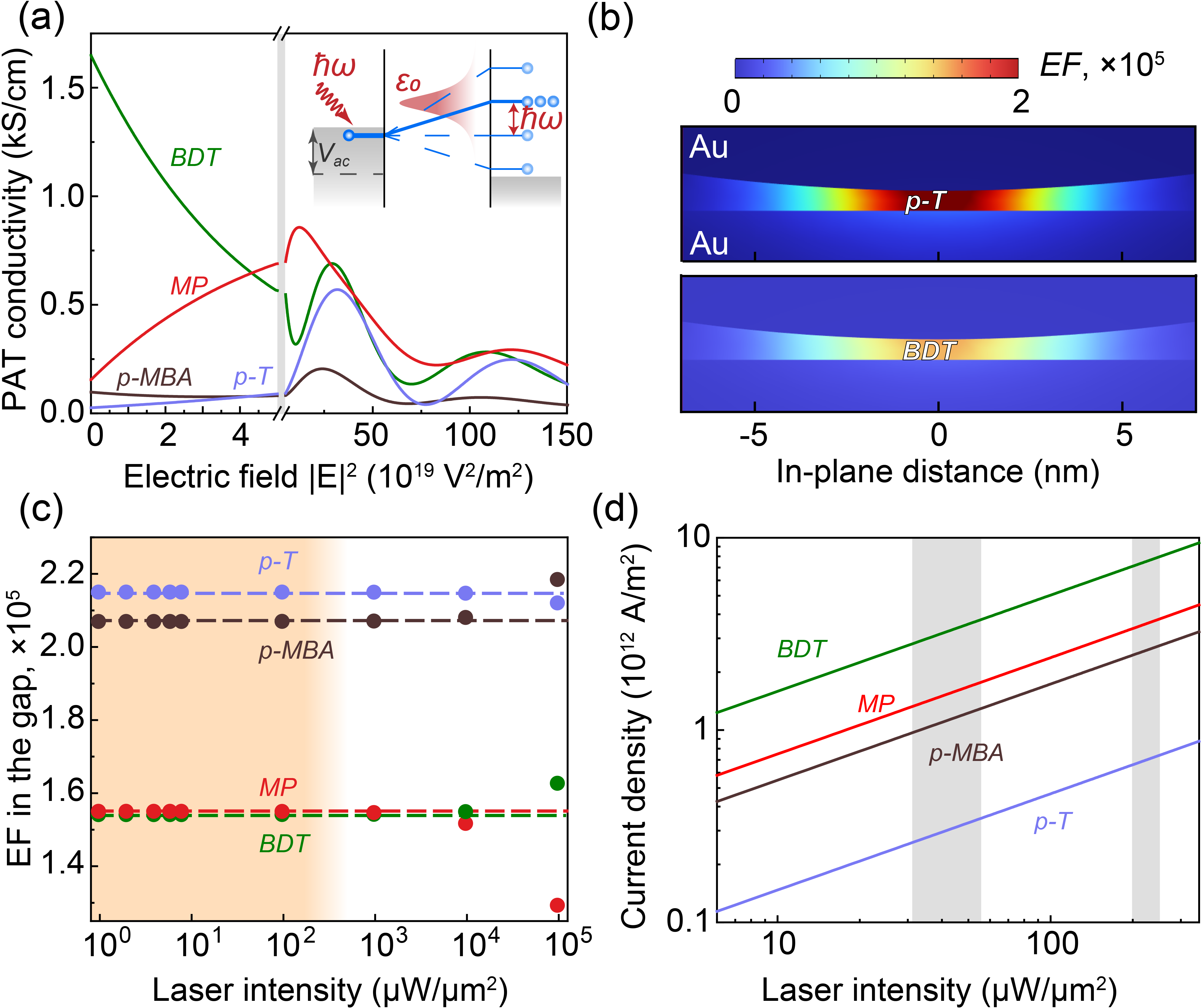}
\caption{{\bf Photo-assisted tunneling conductivity in molecular nanojunctions.} 
(a) Calculated dependence of PAT conductivity of the four studied molecules on the applied electric field. The inset illustrates the energy diagram for PAT.
(b) Simulated spatial distribution of the near-field enhancement factor $EF$ in the p-T (top) and BDT (bottom) molecular nanojunctions. $EF$ is obtained as the square of the electric field $|E|^2$ normalized to the incident light intensity $|E_0|^2$.
(c) The dependence of the enhancement factor of the electric field intensity $EF$ in the NPoM gap on the intensity of the resonant (1.58~eV) excitation, calculated within the PAT formalism. The dashed lines illustrate linear regimes and serve as a guide to the eyes. The orange shaded area indicates the intensity range accessible in the experiments.
(d) The simulated dependence of the peak PAT current densities through the molecular nanojunctions on the illumination intensity. The gray shaded regions highlight the nonlinear regime observed in the experimental data for BDT and p-MBA nanojunctions. 
}
\label{fig:pat}
\end{figure}

Next, we employ the PAT formalism~\cite{TienGordonPR1963} to calculate the electric field-dependent conductance. In short, the conductance of a NPoM molecular junction can be described as
\begin{equation}
    G(\alpha,\omega)=G_{0}\sum^{\infty}_{l=-\infty} J^2_l(eV_{ac}/\hbar\omega) T(\mathcal{E}_F+l\hbar\omega),
\label{eq:conduction}
\end{equation}
where $G_0=2e^2/h$, $\mathcal{E}_F$ is the Fermi energy of the metal, $J_l$ are $l$-th-order Bessel functions of the first kind ($l$ is an integer), and $T(\mathcal{E}_F+l\hbar\omega)$ is the electron transmittance calculated as above. The dimensionless parameter $eV_{ac}/\hbar\omega$ characterizes the optical excitation, where $\hbar\omega=1.58$~eV is the photon energy and $V_{ac}$ is  the electric field-induced voltage across the molecule. The PAT formalism predicts significant conductivity variations at elevated electric fields (Fig. \ref{fig:pat}a).

We embedded the numerically obtained conductivity $\sigma_{\rm PAT}$ (specific conductance) into a full-wave electromagnetic simulation (COMSOL Multiphysics) and calculated the near-field distribution in plasmonic molecular nanojunctions (Fig. \ref{fig:pat}b; see Supplementary Material- Electromagnetic simulations based on finite element method for details). 
High molecular conductance can suppress the plasmonic near-field enhancement\cite{cui2018molecular}. Indeed, NPoM with the least conductive molecule (p-T) demonstrates the strongest electric field in the gap. This reduction of the plasmonic near-field enhancement hampers the nonlinear-optical efficiency in conductive nanojunctions, consistent with our observations: in the low intensity regime ($\leqslant 30$~$\mu$W/$\mu$m$^{2}$), p-T nanojunctions demonstrate the strongest Raman response.

To quantify nonlinear PAT-driven corrections to the near-field intensity enhancement factor $EF\equiv|E|^2/|E_0|^2$, we repeated the calculations for a series of incident laser intensities at the LSP resonant 1.58~eV photon energy (785~nm wavelength). Remarkably, the results indicate that $EF$ variations in nanojunctions with the studied molecules require laser intensities several orders of magnitude higher than those used in the experiments (Fig. \ref{fig:pat}c). This rules out plasmonic ($EF$ variations-driven) origins of the observed Raman nonlinearity; i.e., within the experimentally accessible range ($V_{ac}\lesssim$ 0.01--1~V) of laser intensities, plasmonic properties of the molecular nanojunctions exhibit no noticeable changes.  
The nonlinearity is also not associated with reaching a current threshold, as is seen from the fact that the calculated PAT current density across the nanojunction $j=\sigma_{\rm PAT} E$ exhibits no resonances, but rather shows a monotonic increase for all molecules (Fig. \ref{fig:pat}d). 
This trend is consistent with the S-SERS enhancement for molecules with low conductance. However, for molecular junctions with high conductivity, the PAT current density cannot predict the nonlinear polarizability increase (Fig.~S15).

\subsection{Inelastic electron processes}

The formalism based on electric field-dependent molecular conductance alone cannot quantitatively account for the observed nonlinearity of the S-SERS response in conductive BDT and p-MBA nanojunctions. Instead, a microscopic description of the quantum conductance in molecular junctions should be incorporated \cite{GalperinJPCM2007, vanderMolenJPCM2010, ThossJCP2018}. In short, the tunneling current across the junction contains inelastic contributions which result in phonon excitations (phonon pumping), thereby enhancing Raman scattering~\cite{JaklevicPRL1966}. Because the incident radiation modifies the tunneling rate~\cite{TienGordonPR1963, ButtikerPRL1982}, nonlinearity in the Raman response can be expected. This effect may be further enhanced in the case of resonant tunneling \cite{HalbritterSurf1982, BendingPRL1985, OzakiJAP1998, YuPRL2004}, when the electronic current proceeds through a localized state in the potential barrier.

\begin{figure*}[t]%
\centering
\includegraphics[width=0.7\textwidth]{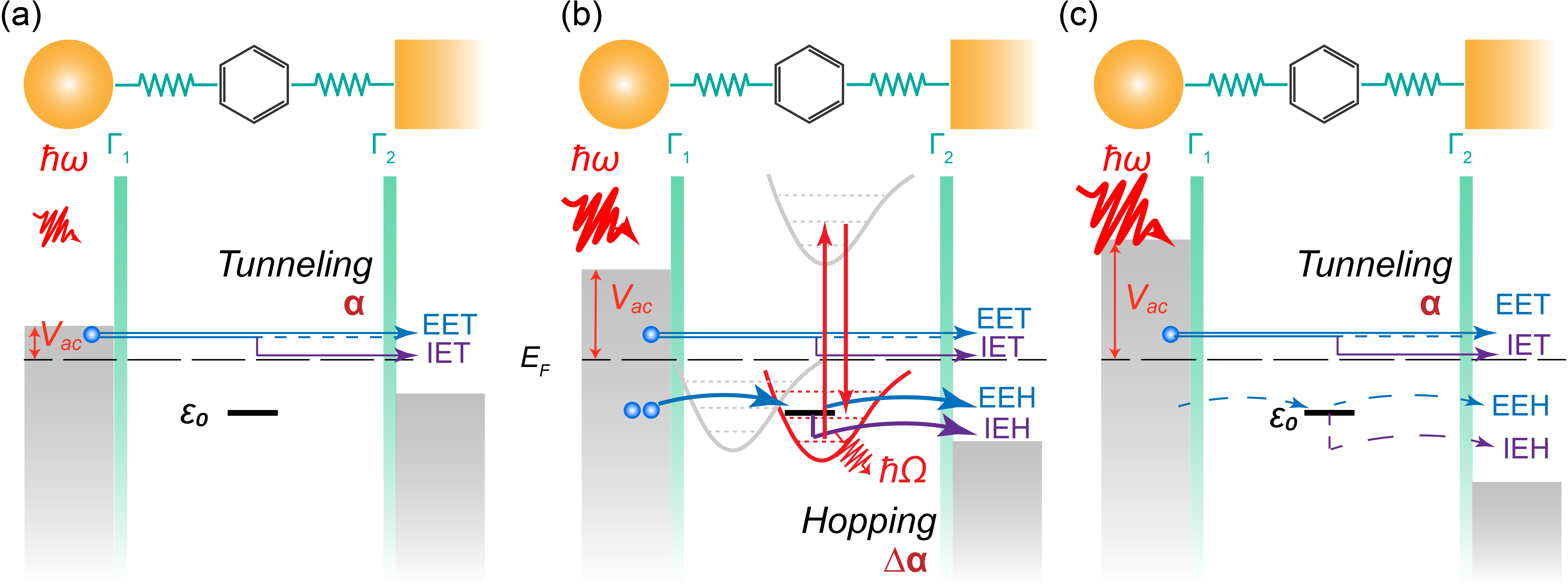}
\caption{{\bf Electron transfer channels in molecular plasmonic nanojunctions.} 
(a--c) Schematics of the Marcus process for the light-field-induced electron hopping conductivity. (a) At low light intensities, only direct tunneling is allowed and Raman polarizability of a molecule is $\alpha$. In this regime, elastic electron tunneling dominates over the inelastic one. (b) When the excitation intensity and thus the electric-field-induced bias voltage $V_{ac}$ is increased, a hopping channel across an energy level on the molecule $\varepsilon_0$ is activated with the transfer rates $\Gamma_{1,2}$. Inelastic electron processes result in the excitation of phonons with an energy $\hbar\Omega$, effectively enabling additional polarizability contribution $\Delta\alpha$. If the molecular electron-phonon coupling in the $\varepsilon_0$ state is strong, inelastic electron hopping (IEH) promotes stimulated phonon emission, and large ratios $\Delta\alpha/\alpha>1$ can be expected (see Fig.~\ref{fig:2}c). (c) As the laser power continues to increase, the light-induced bias voltage further rises, causing the IEH process to weaken until it nearly disappears. The occupied electronic states of plasmonic contacts are represented in gray and the DoS of a BDT molecule is indicated in green.
}
\label{fig:marcus}
\end{figure*}

The Stokes-Raman intensity can be calculated as
\begin{equation}
    R_S(\Omega)=\alpha_{RS}(\Omega)(1+n^{\Omega}_{ph})I,
    \label{eq:stokes}
\end{equation}
where $\alpha_{RS}$ is the effective Raman polarizability, $I$ is the laser intensity, and $n^{\Omega}_{ph}$ is the phonon population at the frequency $\Omega$. This population consists of a thermal phonons $n_{th}=(e^{\hbar\Omega/kT}-1)^{-1}$ and a current-induced phonons $n_c$. Because of inelastic electron scattering, $n_c$ is proportional to the intensity-dependent current and the electron-phonon coupling strength, $n_c\propto j(I) g_{\rm e-ph}$. From Eq.~\ref{eq:stokes}, we then find that $R_S\propto (1+bj(I))I$, where $b$ is a parameter describing the efficiency of the current-induced phonon pumping. Note that phenomenologically, the inelastic electron tunneling can be seen as a contribution to the total Raman polarizability $\Delta\alpha$. Its dependence on the incident intensity is a manifestation of the Raman nonlinearity in conductive plasmonic nanocavities.

To quantitatively assess this inelastic current-induced phonon pumping, we consider a NPoM system as a current junction with an embedded quantum oscillator. The metal particle and the film form electric contacts (Fig. \ref{fig:marcus}). An excitation of the LSPR results in the electric field enhancement in the gap where the molecule is located and thus the applied voltage $V_{ac}$ across the nanojunction.  
This voltage shifts the chemical potential $\mu=\pm eV_{ac}$ in the Fermi-Dirac distribution of the electrons in the metal contacts. In the case of a resonant molecular bridge, the energy levels of the molecule enable electron hopping current~\cite{GalperinPRL2006, LeijnsePRB2008}, which is conventionally discussed within the framework of the Marcus theory \cite{ScalapinoPRL1967} and its extensions~\cite{SowaJPC2018}. Within the Marcus theory, a polaron-transformed Hamiltonian is~\cite{LangFirsov1963} 

\begin{multline}
     \hat{H}=\bar{\varepsilon}a^{\dagger}a+\sum_k \varepsilon_k c^{\dagger}_kc_k +\sum_q \hbar\Omega_q b^{\dagger}_qb_q +\\
     +\sum_k (V_k X^{\dagger}a^{\dagger}c_k+V_k^{*} X a c^{\dagger}_k),
    \label{eq:ham1}   
\end{multline}
where $\bar{\varepsilon}$ is the renormalized energy level of the molecule, $a^{\dagger}$ ($a$) is the electron creation (annihilation) operator, $c_k^{\dagger}$ ($c_k$) is the creation (annihilation) operator for an electron with energy $\varepsilon_k$ in the metal contacts, and $b_q^{\dagger}$ ($b_q$) is the creation (annihilation) operator for phonons with an energy $\hbar\Omega_q$. The last term describes electron-phonon coupling, where $V_k$ is the coupling operator and $X=\sum_q (g_q/\Omega_q)(b_q^{\dagger}-b_q)$ is the displacement operator ($g_q$ is the electron-phonon coupling strength). The current $j\propto\Gamma$ is calculated from the steady-state solution of a quantum master equation (see Supplementary Material and Ref.~\cite{SowaJPC2018}), where molecular couplings to both conductors $\Gamma_1$ and $\Gamma_2$ (illustrated schematically in~Fig. \ref{fig:1}a) are effectively represented as $\Gamma=\Gamma_1\Gamma_2/(\Gamma_1+\Gamma_2)$. The calculated current $j$ can then be incorporated into Eq.~(\ref{eq:stokes}) to determine the corresponding S-SERS intensity.

\begin{figure}[ht]%
\centering
\includegraphics[width=0.48\textwidth]{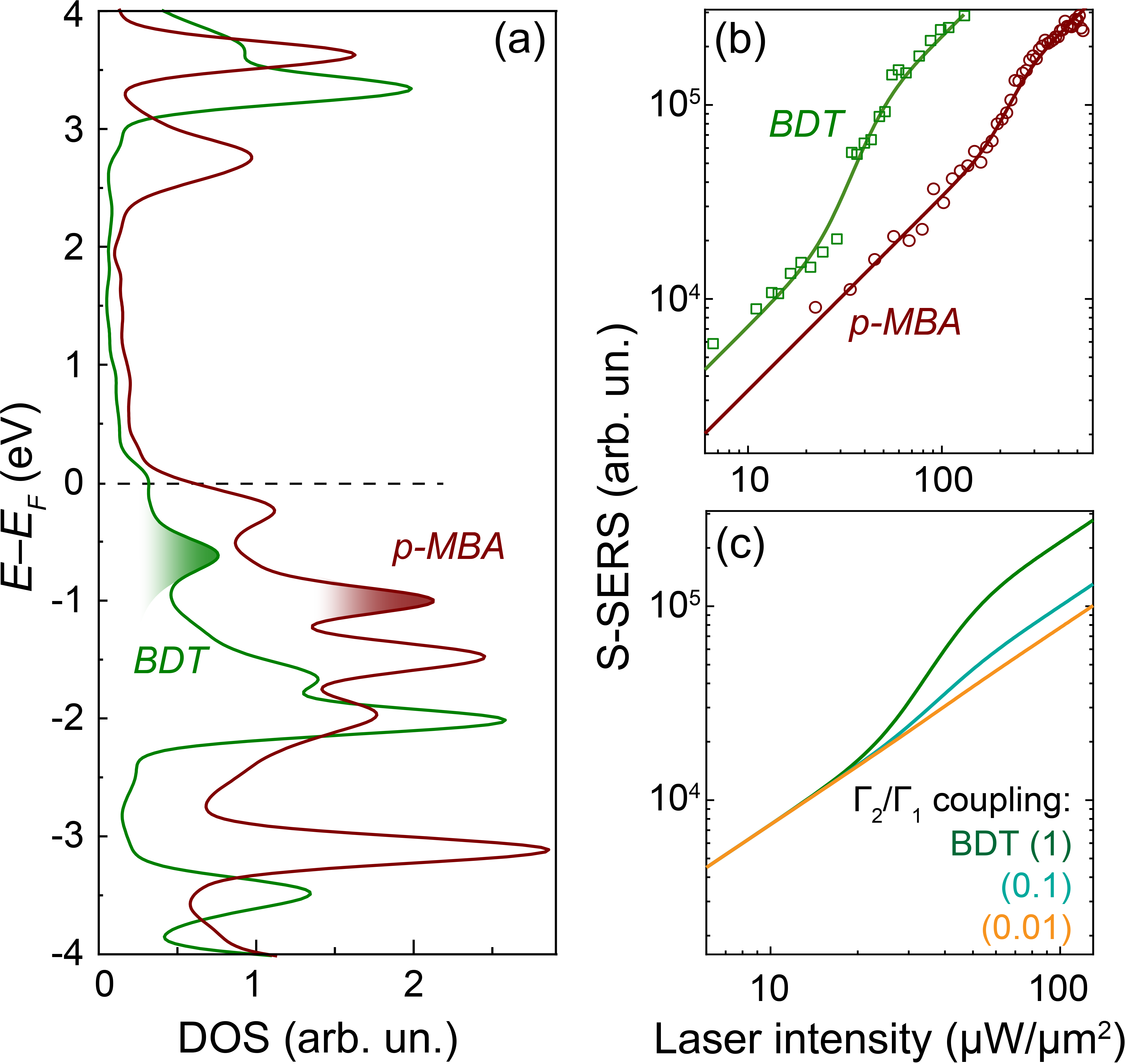}
\caption{{\bf Inelastic Marcus hopping and vibrational pumping in molecular plasmonic nanojunctions.} 
(a) DFT-calculated density of states for the BDT (green) and p-MBA (brown) molecules. The shaded areas indicate resonant electronic states which are used in the Marcus current calculations.
(b) S-SERS intensities (solid lines) calculted within the Marcus model in the nanojunctions with BDT (green) and p-MBA (brown) molecules. The dots show the experimental S-SERS data obtained for the $\sim1070$~cm$^{-1}$ vibrational mode. 
(c) The impact of the molecule-gold coupling strength on SERS nonlinearity. Nonlinear SERS response is observed in the case of symmetric coupling ($\Gamma_1=\Gamma_2=1$) for BDT molecule but not highly asymmetric coupling ($\Gamma_1\gg\Gamma_2$) for p-T molecule). 
}
\label{fig:theory}
\end{figure}

Figure 5b shows a remarkable agreement between the calculated Stokes intensity within the Marcus approach and the experimental Raman spectra for BDT and p-MBA nanojunctions. In the calculations, we used the enhancement factors $EF$ (different for BDT and p-MBA) from Fig. \ref{fig:pat}c and room temperature $kT=25$~meV environment. The energies $\bar{\varepsilon}_{\rm BDT}\approx-0.6$~eV and $\bar{\varepsilon}_{\rm p-MBA}\approx-1.0$~eV were obtained from the DFT calculations of the molecular DOS (see Fig.~\ref{fig:theory}a).  The probability of the BDT-Au electron tunneling averaged over the ensemble of molecules is equal on both sides,  $\Gamma_1=\Gamma_2=1$, whereas for p-MBA we set $\Gamma_2\approx 0.4$. The resonant nature of the current-induced nonlinearity is further emphasized by the simultaneous Stokes response variations of multiple vibrational modes. Together with the CW nature of the excitation, this behavior captured by the developed model makes unlikely alternative explanations related to vibrational pumping \cite{MaherJPCB2006} or impulsive stimulated Raman scattering \cite{RuhmanRPA1987}. Similarly, we exclude Joule heat-induced pumping of thermal phonons \cite{CireraACSNano2022} due to the high energy of the observed vibrational modes and the distinct onset of the nonlinear regime. The tilt angle of molecules on the Au surface may vary slightly due to bonding differences, causing minor fluctuations in gap size beyond the nominal 0.7 nm. This uncertainty, comparable to that of the molecular HOMO level obtained from DFT calculations, does not affect the main conclusions regarding nonlinear Raman responses.

\section*{Discussion}

The Marcus theory for the hopping current is capable of explaining the entire set of experimental observations of phonon excitations in plasmonic molecular nanojunctions. Firstly, non-resonant excitation at the 1.96~eV photon energy does not provide a LSP-induced electric field enhancement in the gap, and $V_{ac}$ is low in this regime. Therefore, the current is also low and the linear dependence of the Raman signal on the intensity is observed throughout the experimentally accessible light intensity range below the damage threshold. 
Secondly, although the LSP-enhanced field in the nanojunctions with the BDT molecules is the weakest (Fig.~\ref{fig:pat}), the nonlinear regime is reached at lower intensities than in the p-MBA nanojunctions. This observation is in a striking contrast with the conventionally assumed mechanisms for the cavity-driven opto-mechanical coupling. Yet, in the p-MBA molecule, the Marcus hopping proceeds through a deeper energy level (Fig. \ref{fig:theory}a), which requires stronger electric fields in the gap and thus higher intensities.  

The conventional approaches to the plasmon-vibrational coupling predict Raman nonlinearities at even lower excitation intensities in the resistive MP and p-T systems ~\cite{cui2018molecular, benz2015nanooptics}. Conversely, we observed no nonlinearity in the studied molecular nanojunctions (cf. Fig. ~\ref{fig:1}d). This apparent contradiction is readily resolved within the Marcus hopping mechanism too. The BDT molecular bridge is symmetric ($\Gamma_1=\Gamma_2$) and thus the hopping contribution (both elastic and inelastic) to the total current $j\propto\Gamma=\Gamma_1\Gamma_2/(\Gamma_1+\Gamma_2)$ is the largest. However, both MP and p-T molecules are asymmetric, meaning that their electronic coupling with one contact is much weaker ($\Gamma_1\gg\Gamma_2$), resulting in a current bottleneck at this contact. In this case, the overall coupling factor is governed by its weakest link, $\Gamma\approx\Gamma_2$, which effectively limits the current and thus the maximal attainable contribution to the Raman (Stokes) response (Eq.~\ref{eq:stokes}). Fig.  \ref{fig:theory}c illustrates the effect of asymmetric coupling: suppression of the resonant (hopping) current cannot be remedied by increasing the voltage bias through applying higher laser intensities. Lastly, the magnitude of nonlinearity 
shown in Fig.~\ref{fig:2}c for various vibrational modes represents the relative efficiency of phonon pumping through Marcus hopping $\Delta\alpha/\alpha$, which is indeed proportional to the mode-specific electron-phonon coupling $g_{e-ph}$. By quantifying the net phonon gain, the mechanism for the opto-mechanical coupling is revealed, illustrating the decisive effect of molecular conductivity.

We note that there are several possible extensions of the Marcus model \cite{EversRMP2020}. It can be further refined by explicitly including phonon-dependent electron-vibrational coupling \cite{SergueevPRL2005}, low temperature ($kT<\hbar\Omega$) corrections or broadening of the hopping energy level \cite{SowaJPC2018}. However, even the basic Marcus formalism is capable of providing valuable insights into the inelastic electron scattering as the physical origin of Raman nonlinearities. This is particularly important in light of recent discussions of opto-mechanical coupling effects in nanoplasmonic systems \cite{RoelliNatNano2016,SchmidtACSNano2016,LombardiPRX2018,XomalisScience2021,JakobNatComm2023}. Despite the apparent phenomenological similarity, our findings indicate a distinct microscopic mechanism behind the Raman nonlinear regime where the inelastic electron current is dominated by Marcus hopping. Thus, a fully quantitative description can be obtained through the calculations of the molecular phonon spectral density and its coupling to the electron dynamics, as shown above. The model presented in this work can be further tested and refined through IET experiments at low temperatures, where more molecular vibrational modes might be observed directly in the derivative of the electric characteristics.

To further elucidate the origins of the nonlinear increase in SERS intensity, we investigated the power dependence of S-SERS and AS-SERS for the BDT molecule in plasmonic nanocavities (Fig. 6). 
The SERS spectra reveal a pronounced increase of the AS intensity at 1070 cm$^{-1}$ and 1558 cm$^{-1}$ peaks with the increase of the excitation laser intensity.  The integrated intensities of the S-SERS and AS-SERS peak for vibrational modes at 1070 cm$^{-1}$ and 1558 cm$^{-1}$ exhibit a non-linear intensity dependence, with a characteristic S-shape. Phonon populations, derived from the intensity ratio $I_{AS}/I_{S}$ (Fig. S20), demonstrate a substantial increase in the nonlinear regime (Fig. 6d-e). These observations further confirm that the characteristic S-shaped profile of the Stokes intensity curve arises predominantly from an increase in phonon population driven by stimulated phonon scattering, rather than solely from modifications in the Raman cross-section \cite{crampton2018junction}. In contrast, variations in Raman activity alone, such as those induced by changes in plasmonic mode volume or molecular orientation, are insufficient to account for the pronounced enhancement observed in anti-Stokes intensity.  

\begin{figure*}[t]%
\centering
\includegraphics[width=1\textwidth]{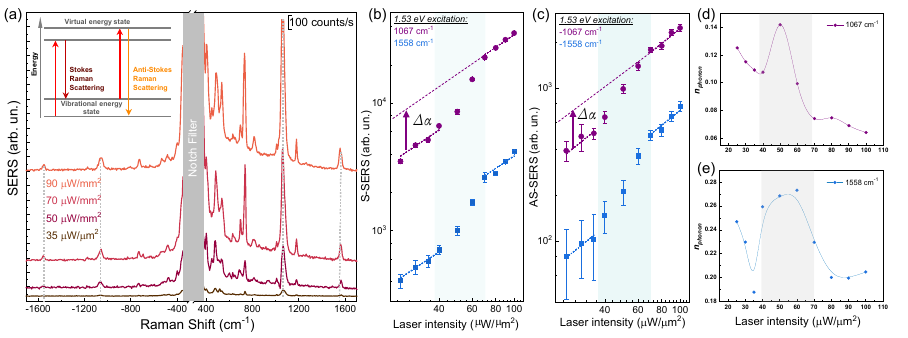}
\caption{{\bf Power dependence of Stokes and AS-SERS in plasmonic nanocavities with BDT molecules.} 
(a) AS- and S-SERS spectra under the excitation at 1.53 eV (808 nm) for laser intensity ranging from 20 to 100 $\mu W/\mu m^2$. (b,c) Excitation intensity dependence of the integrated (b) S-SERS and (c) AS-SERS intensity for vibrational modes $\hbar\Omega\sim$1070 cm$^{-1}$ and 1558 cm$^{-1}$, measured using NPoM structures.
(d-e) Phonon populations calculated based on the $I_{AS}/I_S$ ratio. In the nonlinear regime, the phonon population is significantly elevated. These findings indicate that the enhancement of AS-SERS primarily arises from an increase in vibrational occupancy through a phonon pumping process, rather than solely a change in Raman activity.
}
\label{fig:AS_SERS}
\end{figure*}
\newpage

\noindent Reversibility of the observed nonlinear behaviour allows us to exclude alternative explanations such as redox processes where charge transfer onto the molecule modulates the Raman cross-section (Figs. S16 and S19).  Consequently, the observed intensity dependence of S-SERS confirms the mechanism of enhanced electron-phonon coupling within the Marcus hopping current framework, underlining the intricate interplay between molecular conductance and Raman scattering phenomena. Future investigations, including temperature-dependent studies, might provide quantitative insights into thermal contributions to vibrational occupancy, thereby deepening our understanding of the underlying physical processes. Excitation power cycling experiments further confirm the reversibility and stability of the nonlinear behaviour, demonstrating that the nonlinear SERS response remains consistent under repeated excitation and not related to irreversible modifications of molecules and/or nanostructures.

\vspace{20\baselineskip}

\section{Conclusions}
We observed Raman scattering nonlinearity in conductive molecular nanojunctions at the excitation intensities several orders of magnitude lower than expected in the conventional opto-mechanical approach, with the nonlinear behavior further corroborated by the observation of enhanced anti-Stokes Raman scattering, indicative of increased vibrational occupancy due to phonon pumping. The inelastic electron hopping framework for the description of the observed phonon-pumping was developed by incorporating molecular conductance and electron-phonon coupling into a self-consistent description of the vibrational response of aa molecular electric junction at the nanoscale. The developed framework complements the Raman scattering picture of opto-mechanically driven vibrational modes in plasmonic molecular nanojunctions. The obtained experimental and theoretical results pave a way to further development of molecular nanophotonic and nanoelectronic systems.



\newpage

\begingroup
\let\clearpage\relax 
\onecolumn

\subsection*{Supplementary Material}

\subsection*{Sample preparation}

The nanoparticle-on-mirror (NPoM) structures used in this work consist of a spherical Au nanoparticle and an Au film, separated by a self-assembled molecular layer that forms a molecular junction. As a first step in the fabrication, a homogeneous solution of Au nanoparticles with an average diameter of 80 $\pm$ 6~nm was prepared (472~$\mu$L, $5.8\times 10^{-11}$~mol/L concentration). Subsequently, 50~$\mu$L of the molecular solution was added to the Au nanoparticle solution to achieve a final molecular concentration of 0.1~mM. This process was carried out under intense ultrasound conditions for 30~s, ensuring a precisely controlled addition rate of the molecular solution. The molecules, including 1,4-benzenedithiol (BDT, purchased from Sigma) and other thiol derivatives such as p-MBA, MP, and p-T (purchased from Aladdin), were dissolved in ethanol prior to use. The Raman scattering spectra of the four molecules are shown in Fig. \ref{fig:ramanbulk}.

The number of molecules and the linking duration are the critical factors for coating Au nanospheres with a molecular monolayer. Each molecule occupies the area of approximately 0.22~nm$^2$ on the surface of an Au nanosphere~\cite{benz2015nanooptics}. The number of molecules added is twice that required to form a single molecular layer on each gold nanoparticle. The duration of the molecular linking process on the Au nanosphere surface was approximately 15~min. This timeframe was determined through real-time monitoring of LSP spectral changes during the interaction between molecules and nanospheres~\cite{gandra2012bilayered}. The redshift in the LSP peak of the Au nanoparticle-molecule mixture was recorded using a UV-vis spectrometer. As shown in Fig. \ref{fig:coverage}a,b, the extinction peak of the Au nanospheres exhibits a gradual redshift, stabilizing during the 15-min incubation period with BDT. The initial stage (0–15~min) is attributed to the rapid chemical interaction between the thiol derivative and the Au surface, specifically the formation of Au–S bonds, which drives the self-assembly of molecular monolayers. The subsequent stage is hypothesized to involve the formation of disulfide bonds between the molecules, facilitating the growth of molecular multilayers \cite{lin2018electron}. 

To verify the formation of the molecular monolayer, transmission electron microscopy (TEM) was used to characterize the Au nanospheres incubated with BDT molecules for 15 min. The TEM image reveals a visible monolayer of the BDT molecules with a thickness of approximately 0.7~nm (Fig. \ref{fig:coverage}c). The nanoparticles used in this study are predominantly spherical with minimal faceting, as confirmed by transmission electron microscopy (Fig. S2). Although nanoparticle faceting may influence the electric field enhancement factor, our study does not aim to quantitatively evaluate Raman intensities. Consequently, we adopted a spherical, non-faceted model in the simulations, as faceting does not significantly affect the nonlinear behavior of the Raman polarizability observed in the experiments.

After the incubation, the nanoparticles were thoroughly washed via centrifugation and redispersed in ethanol before being drop-casted onto the Au film. The Au films with a thickness of approximately 200~nm were fabricated using the template-stripping method at a deposition rate of 1.5~Å/s. Initially, a 200 nm-thick Au film is deposited using magnetron sputtering onto an ultra-smooth silicon substrate which has undergone a rigorous cleaning process. This includes sequential ultrasonic cleaning with acetone, isopropanol, and deionized water, followed by plasma cleaning to remove organic residues and impurities. Subsequently, ultraviolet (UV)-curable adhesive is applied to the surface of the Au film, and a pre-cleaned SiO$_2$ substrate is carefully placed onto the adhesive. Under the gravitational force of the SiO$_2$ substrate, the adhesive spreads uniformly between the Au film and the substrate. The assembly is then exposed to UV irradiation to cure the adhesive, securely bonding the SiO$_2$ substrate to the Au film. The substrate is then detached, exposing the Au film surface that was in contact with the ultra-smooth silicon. This approach ensures that the Au film remains isolated from air prior to use, preventing oxidation or contamination by impurities. The Au film is typically used for the nanoparticle drop-casting within 24 hours of preparation to ensure surface cleanliness. High-quality NPoM molecular junctions were subsequently produced by air-drying the samples for 6–10 hours at room temperature. 

\subsection*{Dark field spectroscopy}
The molecular coverage of the Au nanoparticles was further verified with the dark field spectroscopy. The optical dark-field imaging and spectroscopic studies were performed with a customized Olympus BX51 microscope. A 100$\times$ dark-field objective (NA = 0.8) was used to focus unpolarized white light from an incandescent lamp onto the sample.

For the sample, incubated for 90~min, we observed two large groups of spectra where the gap plasmon mode was primarily located at the wavelengths of 780~nm and 750~nm (Fig. \ref{fig:coverage}d). This bimodal distribution is attributed to the monolayer and bilayer molecular coverage on the nanoparticle surface, respectively. These results highlight the critical importance of controlling the incubation time between the molecules and Au nanospheres to achieve monolayer molecular coverage. 
The scattering spectra of the  samples featuring various molecules (BDT, p-MBA, MP and p-T), incubated for 15 min, together with the histogrames of the sample-to-sample variation of LSP wavelengths for various NPoM systems are shown in Fig. \ref{fig:darkfield}.  Compared to the samples incubated for 90~min (Fig. \ref{fig:coverage}d), the resonant wavelengths exhibit a narrower distribution after reducing the incubation time to 15~min. This demonstrates that precise control of the incubation time is essential for achieving uniform monolayer coverage and optimised resonance properties in NPoM systems.

\subsection*{Raman Spectroscopy}

The Raman spectra and images were obtained using a WITec alpha300R confocal Raman microscope (WITec GmbH) equipped with 785~nm (1.58~eV) and 633~nm (1.96~eV) diode lasers. The excitation laser beam was focused through a 100$\times$ objective (NA = 0.9) into a diffraction-limited spot of about 1~$\mu$m in diameter (Fig. \ref{fig:expraman}). The Raman-scattered radiation was collected by the same objective and then dispersed by a high-resolution grating of 600 grooves/mm (UHTS 600). The acquisition time for each spectrum was 20~s.

Having characterised approximately 20 nanoparticles with dark-field microscopy, we selected those with similar resonance response for Raman spectroscopy. The SERS spectra of NPOMs with different molecules were measured under low-power excitation $\approx$1~$\mu$W with integration time of 60~s (Fig. \ref{fig:lowpower}). The particles exhibiting similar S-SERS response were subsequently selected for further investigation.

\subsection*{Opto-mechanical formalism}
We calculated the phonon population $n_{ph}$ (Fig. \ref{fig:optomech}c) and variations of the vibrational linewidth (Fig. \ref{fig:optomech}d) as a function of the incident laser intensity within the opto-mechanical formalism~\cite{AspelmeyerRMP2014, schmidt2017linking}. The calculations predict that significant changes in vibrational damping only emerge at the laser intensities much higher than those employed in our experiments.
Indeed, the steady-state phonon population is given by
\begin{equation}
    n_{ph}=\dfrac{\gamma_m}{\gamma_m+\Gamma_{\rm opt}}n_{th}+\dfrac{\Gamma_+}{\gamma_m+\Gamma_{\rm opt}} \, .
\end{equation}
Here, $\gamma_m$ is the intrinsic vibrational mode damping, and $n_{th}$ denotes the thermal population of phonons. At room temperature T = 300~K, for the high-energy vibrational mode at 1558~cm$^{-1}$ we obtain $n_{th} \approx$5$\times$10$^{-4}$. The Stokes ($\Gamma_+$) and anti-Stokes ($\Gamma_-$) cavity-assisted transition rates emphasize the role of the plasmonic cavity \footnote{Here, and only in this subsection, we use $\Gamma$ as an opto-mechanical contribution to the decay rate of vibrational modes, to retain consistency with other publications.}. The opto-mechanical damping, in turn, is given by \cite{schmidt2017linking}
\begin{equation}
    \Gamma_{\rm opt}=\Gamma_{-}-\Gamma_{+}=g_{0}^2\left|\alpha\right|^{2}\kappa \left[\dfrac{1}{(\Delta-\omega_m)^2+(\kappa/2)^2}-\dfrac{1}{(\Delta+\omega_m)^2+(\kappa/2)^2}\right] \, ,
\end{equation}
where $\hbar g_0$ is the single-photon opto-mechanical coupling strength, $\kappa$ is the plasmon damping, and the coherent amplitude $\alpha$ in typical SERS conditions at optical intensity $I$ can be found as 
\begin{equation}
   |\alpha|^2=\frac{6\pi c^2\kappa\eta}{\kappa^2+4\Delta^2}\frac{I}{\hbar\omega_{\rm LSPR}^3} \, 
\end{equation}
with $\eta<1$ being the radiative yield.
This equation indicates how a plasmonic cavity can either enhance (for $\Gamma_{\text{\rm opt}}>0$) or suppress (for $\Gamma_{\text{\rm opt}}<0$) the effective decay rate $\Gamma_{\text{\rm eff}}\equiv \gamma_m+\Gamma_{\text{\rm opt}}$ of the vibrational mode. In the calculations, we used $\hbar\gamma_m=$12~cm$^{-1}\approx$ 1.5~meV from the fitting, whereas for $\hbar g_0\approx$0.38~meV, 
we used the value from Ref.~\cite{LombardiPRX2018}. Notably, we did not observe a significant increase in the vibrational linewidth $\gamma_m$ when comparing SERS results from NPoM with those obtained in molecular solutions (Fig. \ref{fig:ramanbulk}). Estimating the number of molecules in the laser spot contributing to the total Raman response $N_m\approx$ 10$\times$10=100, we get $\hbar g_{\rm N=100} = \hbar g_0 \sqrt{N_m} \approx$ 4~meV. For the plasmon damping, we take $\kappa=$100~meV, accounting for the large width of the LSP resonance in the extinction spectrum. The simulations based on this conventional opto-mechanical model for the dynamical back-action indicate that significant changes in vibrational damping will only occur at laser intensities far beyond those used in our experiments (Fig. \ref{fig:optomech}d).

\subsection*{DFT calculations}
Calculations of the densities of states (DOS) and electronic transmittance $T$ of the four molecules used in this work were performed within the non-equilibrium Green function (NEGF) approach \cite{datta1997electronic, toher2008effects, soler2002siesta} implemented in the package tranSIESTA from SIESTA \cite{soler2002siesta,brandbyge2002density}. Our model is directly related to the realistic molecular junctions connected to Au (111) surfaces by doing geometry relaxation in the first step, where the exchange–correlation functional is described by Perdew–Burke–Ernzerhof (PBE) generalized gradient approximation (GGA) \cite{kresse1996efficient,perdew1996generalized,blochl1994projector}. The relaxed atomic structure is displayed in Fig. \ref{fig:dfttrans}a. 

Generally, the NEGF scheme partitions the Au-molecule-Au nanojunction into three regions: the two conductors and the middle region (scattering region, SR). The latter includes the molecule and a part of the conductors, described by a Hamiltonian $H_{\mathrm{s}}$. The non-equilibrium Green function $G$ is then constructed as 
\begin{equation}
G=\lim _{\eta \rightarrow 0}\left[\varepsilon-H_{\mathrm{s}}-\Sigma_1-\Sigma_2\right]^{-1} \, ,
\end{equation}
where $\varepsilon=E+i \eta$ ( $\eta$ is an arbitrarily small positive real number) is the complex energy, and $\Sigma_{1,2}$ is the self-energies for the conductors on two ends of the molecule. The Green function $G$ allows us to calculate the SR density matrix from the non-equilibrium charge density in a self-consistent manner once the density matrix converges (to a tolerance of 1$\times$10$^{-5}$ a.u.). Finally, the converged Green function was used to obtain the electron transmittance $T$: 
\begin{equation}
T= \operatorname{Tr}\left[G \Lambda_1 G^{\dagger} \Lambda_2\right] \, ,
\end{equation}
where $\Lambda_{1,2}=i\left[\Sigma_{1,2}-\Sigma_{1,2}^{\dagger}\right]$. Fig.  \ref{fig:dfttrans}b shows the calculated electron transmittance (the Fermi energy is set to 0). The most energetically favorable adsorption site for each molecule in our calculation is the hollow site, in agreement with the previous reports \cite{gronbeck2000thiols,tachibana2002sulfur,nara2004density}.
 
In the electron transmittance spectrum, we anticipate the two peaks which are the closest to the Fermi level referring to electrons from HOMO and LUMO. Hence, we also provide a direct comparison with the local density of states (LDOS) spectrum of each molecule. The HOMO and LUMO peaks demonstrate a good correlation with the relative peak in the transmission spectrum. For a better comparison with the experiment, we shift the empty states to match the gap between HOMO and LUMO with the value obtained from the experimental absorption spectrum. The correction is based on the fact that the quasiparticle wavefunction is very similar to those from the Perdew–Burke–Ernzerhof generalized gradient approximation, with the main difference being a step-like correction at the HOMO-LUMO gap \cite{hybertsen1986electron}. The obtained molecular LDOS was verified to produce optical absorption spectra consistent with the experimental observations ( Fig. \ref{fig:dftdos}). 

\subsection*{Photo-assisted tunneling current }
Photo-assisted tunneling current (PAT) describes the modulation of tunneling currents in nanoscale systems due to interaction with an optical field. According to the PAT model \cite{TienGordonPR1963}, the conductance of a NPoM molecular junction can be described as
\begin{equation}
    G_{dc}(\omega)=G_{0}\sum^{\infty}_{l=-\infty} J^2_l(eV_{ac}/\hbar\omega) T(E_F+l\hbar\omega) \, ,
\label{eq:conduction2}
\end{equation}
where $E_F$ is the Fermi energy of the metal, $J_l$ are the $l$-th-order Bessel functions of the first kind with $l$ being the index of the sideband, $T(E_F+l\hbar\omega)$ is the zero-bias equilibrium electron transmittance at energy $E_F+l\hbar\omega$, and $ eV_{ac}/\hbar\omega $ is determined by the induced ac voltage. The zero-bias equilibrium transmission spectra of the molecules used in our experiments are calculated by the DFT simulations. With the induced $V_{ac}$, the conductivity variations with the optical intensity can be found. Because the electric field in the gap is nearly homogeneous and normal to the film plane, the induced ac voltage is $V_{ac}\approx dE$, where $d$ is the gap width. The induced electric field $E$ can be determined from the Helmholtz equation $\nabla\times(\nabla\times E)-k_{0}^{2}\varepsilon_{r}E$=0. The relationship between the conductance and conductivity is $\sigma_{gap}=G_{dc}(\omega)g_0/s_0$, where $g_0$ and $s_0$ are the length and the cross-section of an individual molecule in the gap, respectively. The current density in the gap is then obtained as 
\begin{equation}
      j_{PAT}=(\vec{n}\cdot\sigma_{gap}+i\omega_0\omega_d)E \, ,
\end{equation}
where $\varepsilon_{d}$ is the dielectric function of the molecule, which was optimized to ensure a good correspondence between the calculated and experimental dark field scattering spectra of the respective NPoM systems. 

\subsection*{Electromagnetic simulations based on finite element method}
We used a full-wave electromagnetic simulation combined with the PAT model to compute the near-field of NPoM structures containing four molecules. The dielectric constant of Au was obtained from the empirical data~\cite{johnson1972optical}. An obliquely incident plane wave was employed to excite the entire structure. The scattering spectra of NPoM were acquired by collecting the upward scattering power flow within a solid angle of 103$^o$, corresponding to an NA value of 0.8 for the objective used in the experiment. The mesh refinement in the simulation model ensured convergence of the computational results.
The incident light power used in the experiment is defined as
\begin{equation}
    P(\lambda^{'})= \frac{P_0}{\pi r^2}.\frac{I(\lambda^{'})}{\int_{\lambda^{'}_2}^{\lambda^{'}_1}I(\lambda^{'})d\lambda^{'}} \, ,
\end{equation}
where $P_0$ represent the total incident power, and $r$=0.16$\lambda^{'}/NA$ denotes the radius of the spot on the sample, with NA being the numerical aperture. The term $I(\lambda^{'})$ corresponds to the measured light intensity spectrum (Fig. \ref{fig:comsolmodel}a).
The PAT current derived above was then incorporated into the Maxwell's equations ($\nabla \times \mathbf{H} = \mathbf{J} + \frac{\partial \mathbf{D}}{\partial t}$) as an additional current term. By including the PAT current in the gap, the molecular conductance is effectively accounted for in the simulations. 

\subsection*{Marcus current calculations}
The Marcus hopping current is obtained by solving a second-order quantum master equation (QME) for the density matrix $\rho$ in the steady state~\cite{SowaJPC2018}.
We begin with calculating the effective bias $V_{ac}$ at various laser intensities $I$ as $V_{ac}=E\times d$, where $E$ is the electric field in the gap and $d=$ 0.7~nm is the gap width. The electric field $E$ was calculated accounting for the plasmonic field enhancement factors obtained from the FDTD simulations. These bias voltages  $V_{ac}$ were then used to symmetrically shift the Fermi-Dirac electron occupation numbers in the contacts: $f_{1,2}(\varepsilon)=(1+e^{-\beta(\varepsilon-eV_{ac}/2)})^{-1}$, where $\beta=1/k_BT$ is the inverse temperature. Although the electron distribution under laser irradiation is non-thermal, in this work we consider a Fermi-Dirac shape for simplicity. Arbitrary electronic distribution can be incorporated into the model, yielding small corrections to the net tunneling current.

In the steady state, the electron transfer rates from the metal contacts to the molecule ($\gamma$) and back ($\bar\gamma$) can be found from setting $d\rho/dt=0$ in the QME. The solution to this equation $\rho_{st}$ takes the following form \cite{SowaJPC2018}:
\begin{equation}
    \rho_{st}=
    \begin{pmatrix}
        \frac{\bar\gamma_{1}+\bar\gamma_{2}}{\gamma_{1}+\gamma_{2}+\bar\gamma_{1}+\bar\gamma_{2}} & 0\\
        0 & \frac{\gamma_{1}+\gamma_{2}}{\gamma_{1}+\gamma_{2}+\bar\gamma_{1}+\bar\gamma_{2}}
    \end{pmatrix},
\end{equation}
where $\gamma_{1,2}$ and $\bar\gamma_{1,2}$ can be found as
\begin{equation}
    \gamma_{1,2}=\frac{1}{\pi}{\rm Re}\,\Gamma_{1,2}\int^{\infty}_{-\infty} d\varepsilon\,f_{1,2}(\varepsilon)\int^{\infty}_0 d\tau\, e^{+i(\varepsilon-\varepsilon_0)\tau}B(\tau) 
\end{equation}
and
\begin{equation}
    \bar\gamma_{1,2}=\frac{1}{\pi}{\rm Re}\,\Gamma_{1,2}\int^{\infty}_{-\infty} d\varepsilon\,[1-f_{1,2}(\varepsilon)]\int^{\infty}_0 d\tau\, e^{-i(\varepsilon-\varepsilon_0)\tau}B(\tau) \, .
\end{equation}
Here, $\Gamma_{1,2}$ denotes the strength of molecule coupling to the metal contacts, and $B$ is the phononic correlation function $B(t-\tau)=\langle X(t)X^{\dagger}(\tau)\rangle$, where $X=\exp\left[ -\sum_q \frac{g_q}{\omega_q}(b_q^{\dagger}-b_q)\right]$ is the displacement operator. For thermalized vibrational modes, $B$ takes the following form:
\begin{equation}
    B(t)=\exp \left(-\sum_q \frac{g_q^2}{\omega_q^2}\times\left[N_q\left(1-e^{i\omega_qt}\right)+\left(N_q+1\right)\left(1-e^{-i\omega_qt}\right)\right]\right) \, ,
\end{equation}
where $N_q = (e^{\beta\omega_q}-1)^{-1}$ is the average population of the vibrational mode $q$ at the inverse temperature $\beta$. Employing the definition of the phonon spectral density $J(\omega)=\sum_q |g_q|^2\times\delta(\omega-\omega_q)$, we obtain
\begin{equation}
    B(t)=\exp\left[\int_0^{\infty}d\omega\frac{J(\omega)}{\omega^2}\left(\coth\left(\frac{\beta\omega}{2}\right)(\cos\omega t -1)-i\sin\omega t \right)\right] \, .
    \label{corrfunc}
\end{equation}
Note that in the absence of the electron-phonon coupling ($J(\omega)=0$, or $B(\tau)=1$), the Landauer-B\"{u}ttiker expression for the (elastic) tunnel current through a single non-interacting energy level can be recovered \cite{ButtikerPRL1982}.

Expanding the trigonometric functions in $B(t)$ (Eq.\ref{corrfunc}) up to the lowest order and introducing reorganization energy $\hbar\lambda=\int d\omega J(\omega)/\omega$, we arrive at
\begin{equation}
    B(t)=\exp\left[-\lambda t^2\beta-i\lambda t\right] \, .   
\end{equation}
This correlation function can be employed in calculating the electron hopping rates $\gamma$ within the Marcus theory:
\begin{equation}
    \gamma_{1,2}=\Gamma_{1,2}\int^{\infty}_{-\infty} d\varepsilon\,f_{1,2}(\varepsilon)\sqrt{\frac{\beta}{4\pi\lambda}}\exp \left[-\frac{\beta\left[\lambda-(\varepsilon-\varepsilon_0)\right]^2}{4\lambda}\right]
\end{equation}
and
\begin{equation}
    \bar\gamma_{1,2}=\Gamma_{1,2}\int^{\infty}_{-\infty} d\varepsilon\,\left[1-f_{1,2}(\varepsilon)\right]\sqrt{\frac{\beta}{4\pi\lambda}}\exp \left[-\frac{\beta\left[\lambda+(\varepsilon-\varepsilon_0)\right]^2}{4\lambda}\right] \, .
\end{equation}
The dependence on the incident laser intensity $I$ in these formulae is hidden in the Fermi-Dirac occupation numbers $f_{1,2}$ in the metal contacts, which shift as a function of the effective bias voltage $V_{ac}$. The current can then be found as
\begin{equation}
j=\Gamma (\gamma_1\bar{\gamma_2}-\gamma_2\bar{\gamma_1})(\gamma_1+\gamma_2+\bar{\gamma_1}+\bar{\gamma_2})^{-1} \, ,
\label{eq:marcuscurrent}
\end{equation}
where $\Gamma=\Gamma_1\Gamma_2/(\Gamma_1+\Gamma_2)$ is the effective coupling parameter between the molecule and the two contacts. 

The individual couplings $\Gamma_1$, $\Gamma_2$ are chosen phenomenologically based on the chemical analysis. All four molecules employed in this work feature an $-SH$ group at one end, which enables a chemical bond to $Au$ and thus high electronic conductance. This allowed us to set $\Gamma_1=$1 for all molecules thus reducing the functional dependence of the Marcus inelastic current to the ratio $\Gamma_2/\Gamma_1$. Further, in symmetric BDT molecules, $\Gamma_2/\Gamma_1=$1 can be assumed. In other molecules employed in this work, $\Gamma_2/\Gamma_1<$1 needs to be taken into account, since in those cases, chemical covalent bonds between the molecule and the Au lead cannot be formed, thus increasing the electronic resistance \cite{BurklePRB2012, cui2018molecular}. 

The inelastic current $j$ from Eq.~(\ref{eq:marcuscurrent}) is responsible for the generation of phonon population $n_ph$ or, in other words, gives rise to the Marcus contribution to the Raman polarizability: $\Delta\alpha\propto j$. To calculate the total Raman response as a function of the incident laser intensity, we incorporate this polarizability contribution into the total Raman intensity as $R=(\alpha+\Delta\alpha)I$, where $\alpha$ is the polarizability producing the Raman response at low excitation intensities in the experiment. Note that $\Delta\alpha$ is governed by the electron-phonon coupling which is different for different vibrational modes observed in the experiments. As such, assuming that the low-intensity response is dominated by the spontaneous Raman processes, the ratio $\Delta\alpha/\alpha$ represents the mode-specific electron-phonon coupling in the HOMO electron state of the molecules. The particular shape of the nonlinear feature is governed by the DOS profile and is beyond the scope of this manuscript.


\renewcommand{\thefigure}{S\arabic{figure}}
\setcounter{figure}{0}

\subsection{Supplementary figures}

\begin{figure}[h]%
\centering
\includegraphics[width=0.65\textwidth]{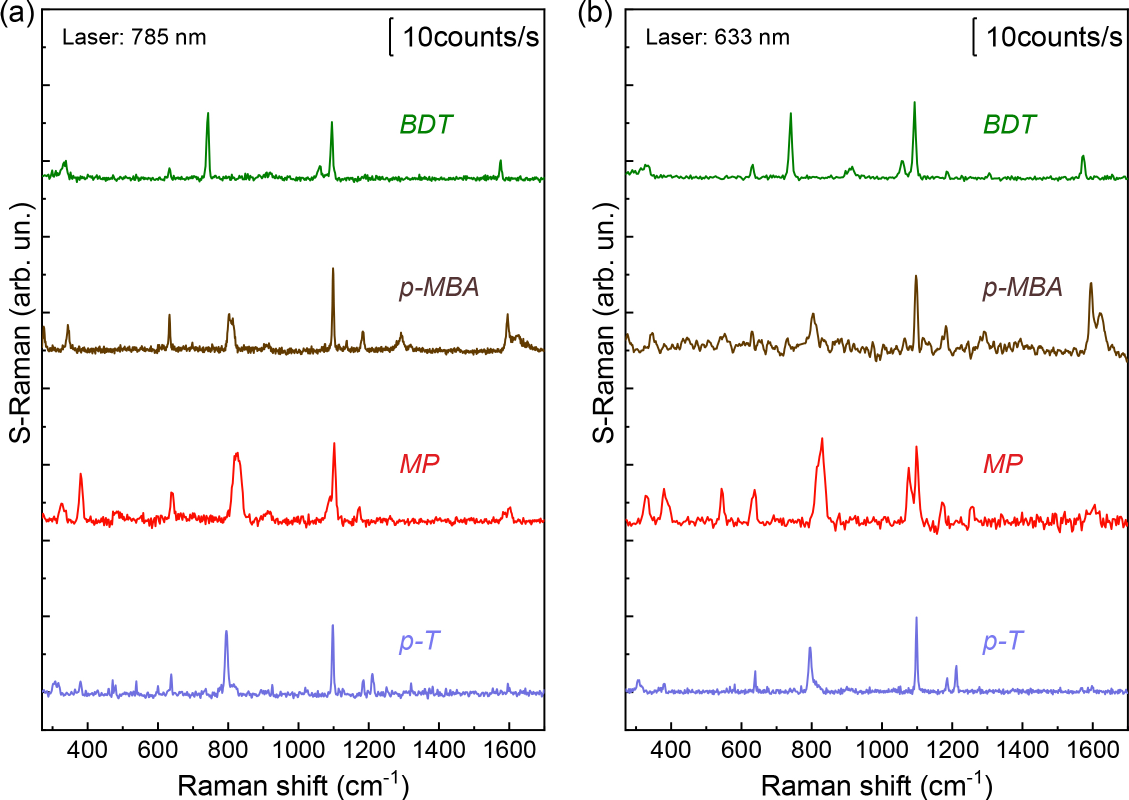}
\caption{{\bf Raman spectroscopy of molecular species.} Stokes-Raman spectra of the the studied molecules measured from a powder. The excitation wavelengths are (a) 785 nm and (b) 633 nm. The incident laser intensity is 7.9 $\mu$W/$\mu$m$^2$.}
\label{fig:ramanbulk}
\end{figure}

\begin{figure}[h]%
\centering
\includegraphics[width=0.65\textwidth]{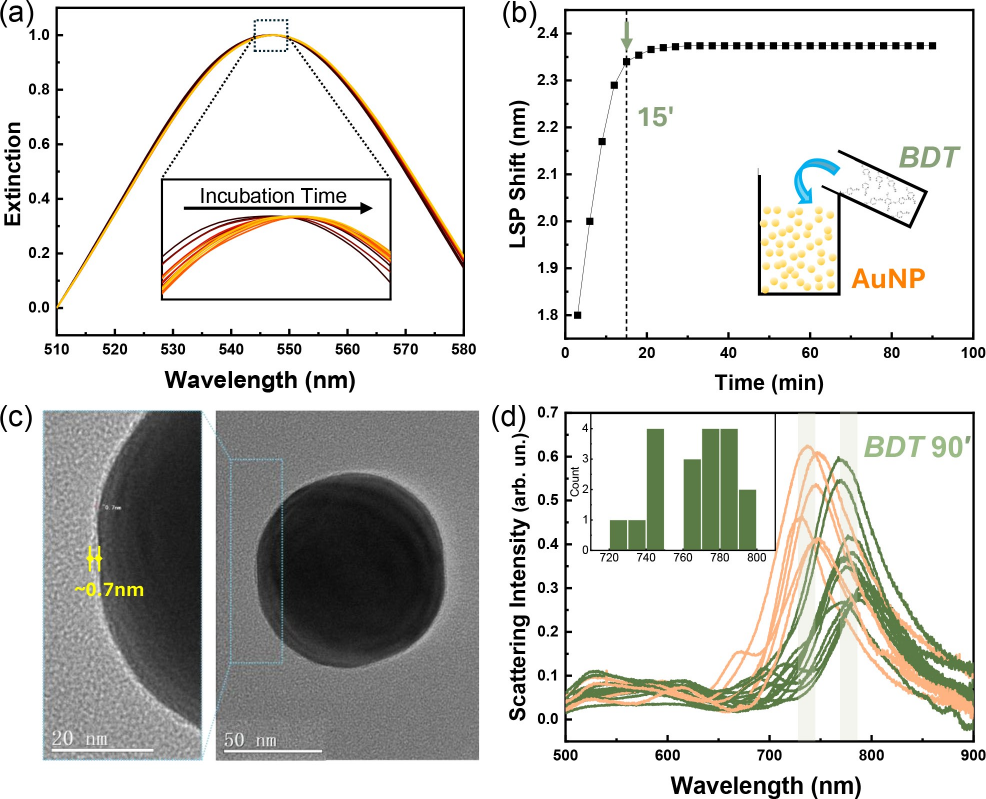}
\caption{{\bf Covering the Au nanoparticles with molecular monolayers.}
(a) Extinction spectra of the Au nanoparticles during the 90 min incubation with BDT molecules.
(b) The LSP shift measured on an Au nanosphere during the 90 min incubation with BDT molecules. The inset shows the experiment schematics. 
(c) TEM images of an Au nanosphere after 15 min incubation with BDT molecules. A molecular monolayer of 0.7~nm thickness can be seen in the zoomed-in image on the left.
(d) Measured single-particle dark-field scattering spectra of multiple NPoMs systems (Au nanosphere with BDT molecules, incubated for 90 min). The gap mode is mainly found around the wavelengths of 780~nm and 750~nm  (gray-shaded areas), corresponding to the monolayer and bilayer molecular coverage of the Au surface, respectively. The inset shows the statistics of the peak wavelengths of the gap mode.
}
\label{fig:coverage}
\end{figure}

\newpage
\begin{figure}[h]%
\centering
\includegraphics[width=0.65\textwidth]{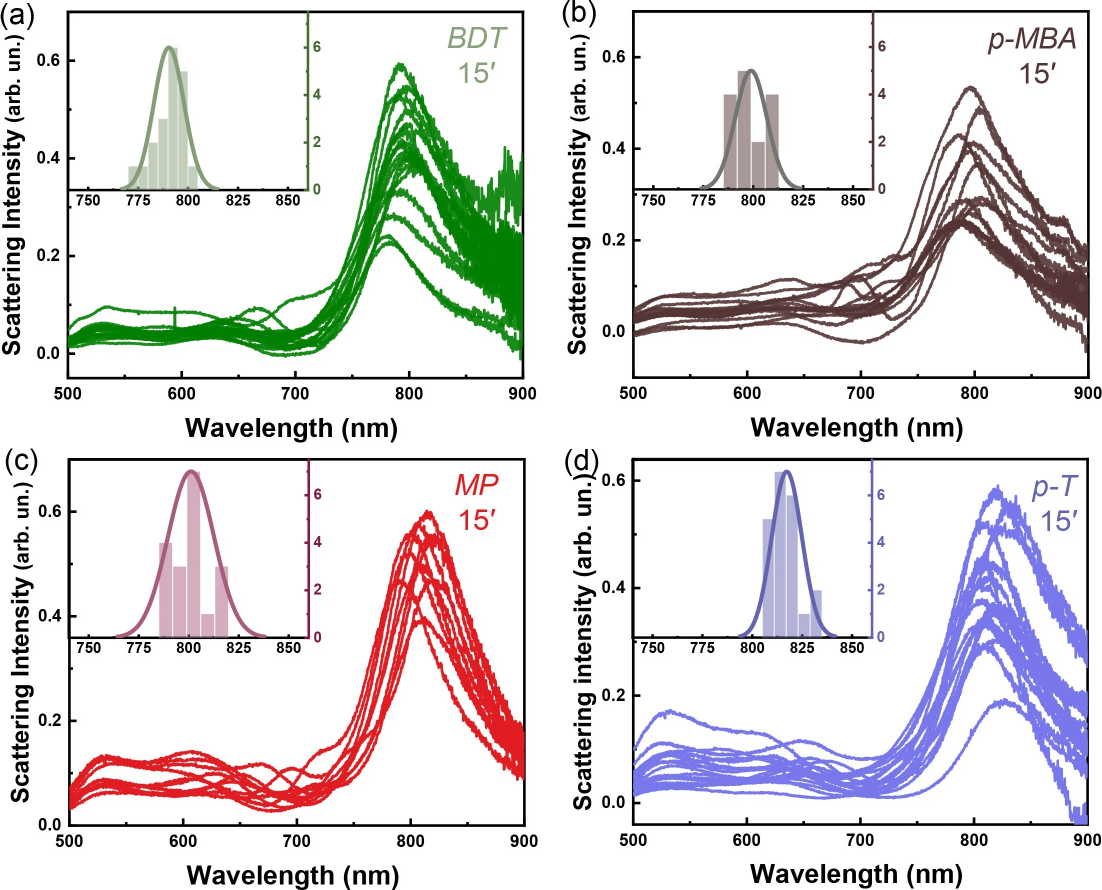}
\caption{{\bf NPoM characterization with dark field spectroscopy.}
Measured single-particle dark-field scattering spectra of the NPoM molecular junctions after 15 min incubation with different molecules: (a) BDT, (b) p-MBA,(c) MP, and (d) p-T. Only monolayer-covered systems are measured. The insets show the statistics of the peak wavelengths of the gap mode for the respective molecular systems with the solid lines representing a fitted Gaussian distribution function.
The averaged wavelengths of the gap-LSP mode in the BDT, p-MBA, MP, and p-T-functionalized NPoMs are: 788$\pm$15~nm, 794$\pm$15~nm, 810$\pm$17~nm and 819$\pm$10~nm, respectively. Note the better match of the excitation wavelegnth (785 nm) with the gap mode peak for the BDT molecule.
}
\label{fig:darkfield}
\end{figure}

\begin{figure}[h!]%
\centering
\includegraphics[width=0.65\textwidth]{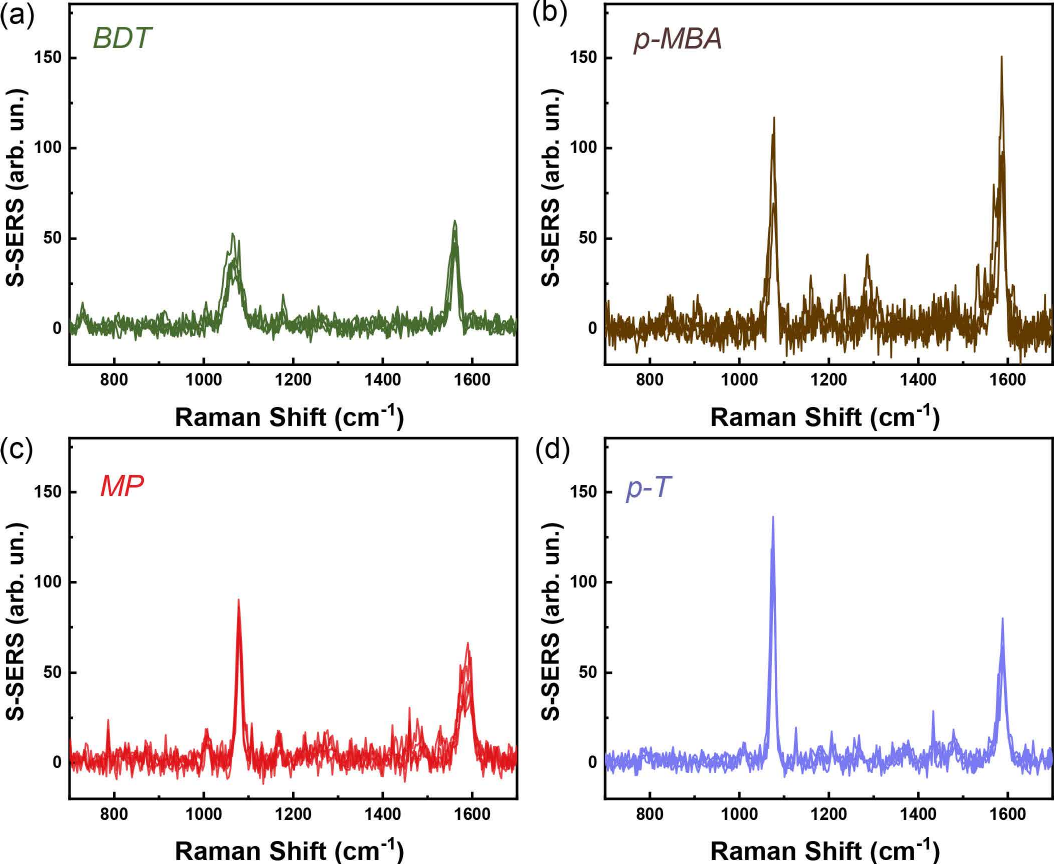}
\caption{S-SERS spectra measured at a low power ($\approx$1~$\mu$W, integration time 60~s). 
Multiple curves in each panel correspond to different individual Au nanoparticles. The NPoM systems demonstrating both strong SERS signals and good stability are selected for the subsequent experiments.
}
\label{fig:lowpower}
\end{figure}
\newpage

\begin{figure}[h]%
\centering
\includegraphics[width=0.65\textwidth]{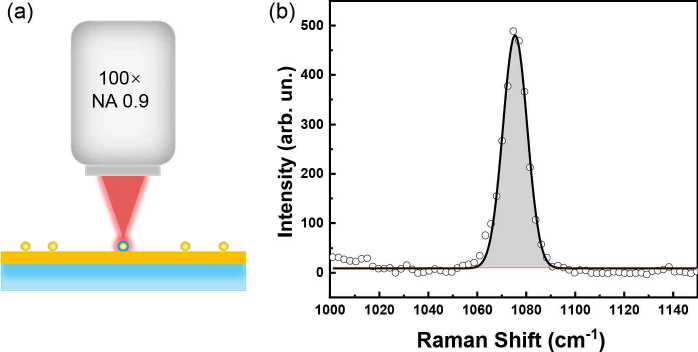}
\caption{(a) Schematic diagram of the single-particle Raman scattering measurements. The excitation light of either 633~nm or 785~nm wavelength is focused with a NA = 0.9 objective. The same objective is used to collect the emitted radiation. (b) S-SERS spectrum taken from BDT NPoM in the vicinity of the $\hbar\Omega\approx$1070~cm$^{-1}$ vibrational mode (open dots). Background subtraction and Gaussian fitting (black line) allow to obtain a S-SERS peak area (gray shaded area).}
\label{fig:expraman}
\end{figure}

\begin{figure}[h]%
\centering
\includegraphics[width=0.4\textwidth]{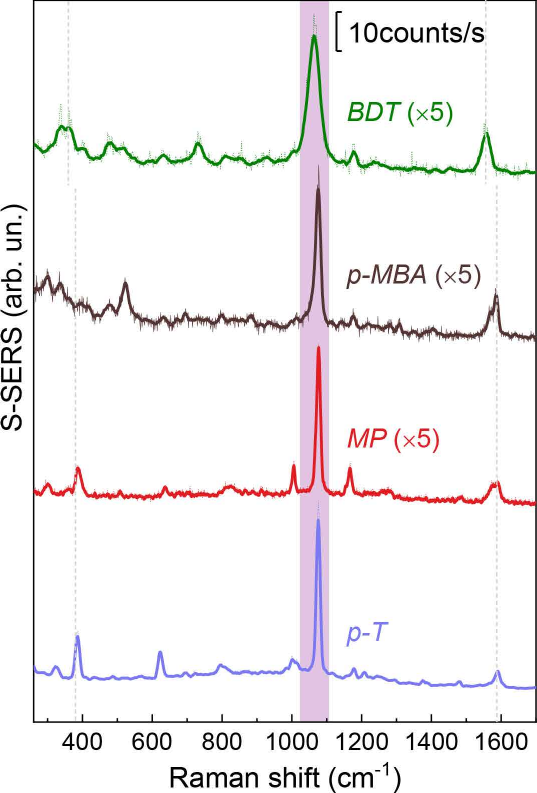}
\caption{Raman spectra from NPoM nanojunctions with different molecules under on-resonant LSP excitation (1.58 eV, 785~nm; cf. Fig.~1b in the main text). The shaded area indicates the prominent $\hbar\Omega\approx$1070~cm$^{-1}$ vibrational mode.}
\label{fig:ramanres}
\end{figure}
\newpage

\begin{figure}[h]%
\centering
\includegraphics[width=0.9\textwidth]{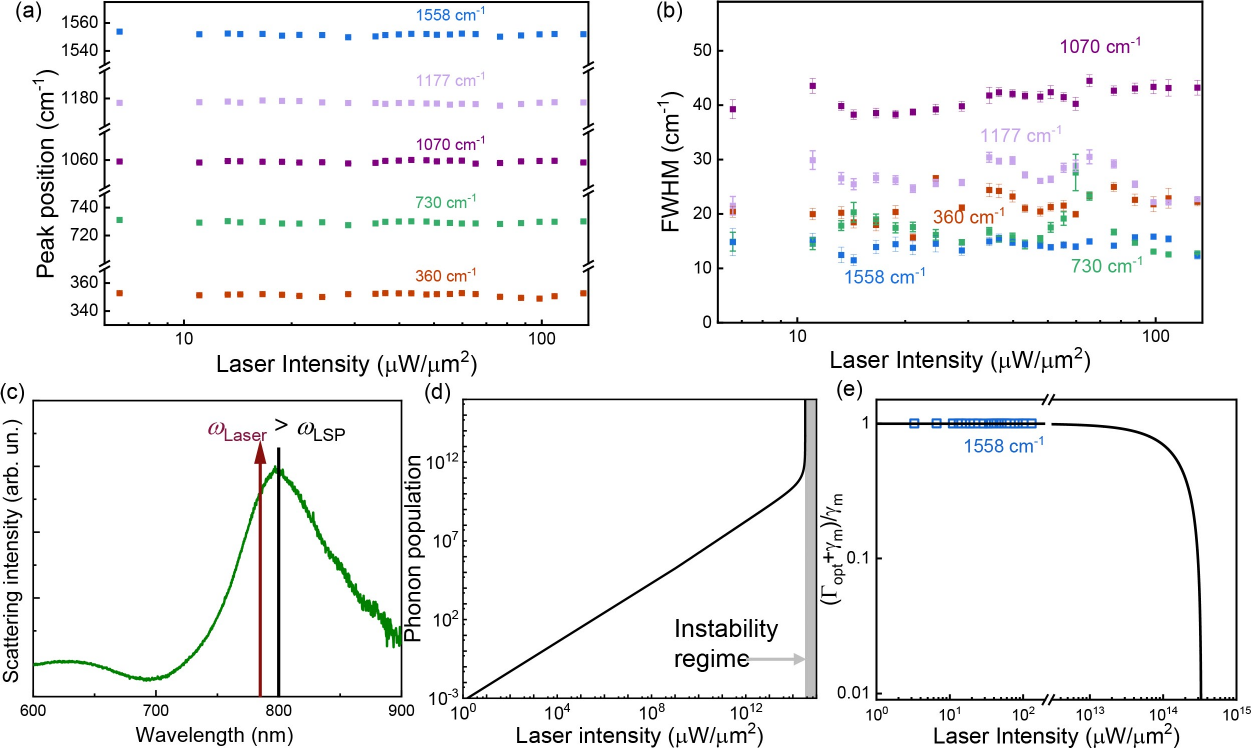}
\caption{{\bf Analysis of the spectral characteristics of the Raman peaks.} Excitation intensity dependence of (a) the peak positions and (b) full widths at half maximum (FWHM) of the five detected S-SERS peaks in the BDT-functionalized NPoMs. No reduction of the linewidth or shift of the mode frequency is observed in the experimentally accessible range of the excitation intensities.
(c) The dark-field scattering spectrum illustrates the position of the excitation wavelength with respect to the LSP resonance. (d) The phonon population and (e) their vibrational decay rate $\Gamma_{\text{eff}}$ calculated in the opto-mechanical dynamical back-action model following Ref.~\cite{LombardiPRX2018}.
The calculated vibrational decay rate is normalized to the low-intensity value $\gamma_{\text{m}}$ at 1558~$\text{cm}^{-1}$. The blue squares in (e) show the FWHM of mode at 1558~$\text{cm}^{-1}$ as obtained from the Lorentzian fitting of the experimental data.
The opto-mechanical dynamical back-action model predicts significant modifications in vibrational damping only at much higher laser intensities than those employed in the experiments. 
} 
\label{fig:optomech}
\end{figure}

\begin{figure}[h]%
\centering
\includegraphics[width=1\textwidth]{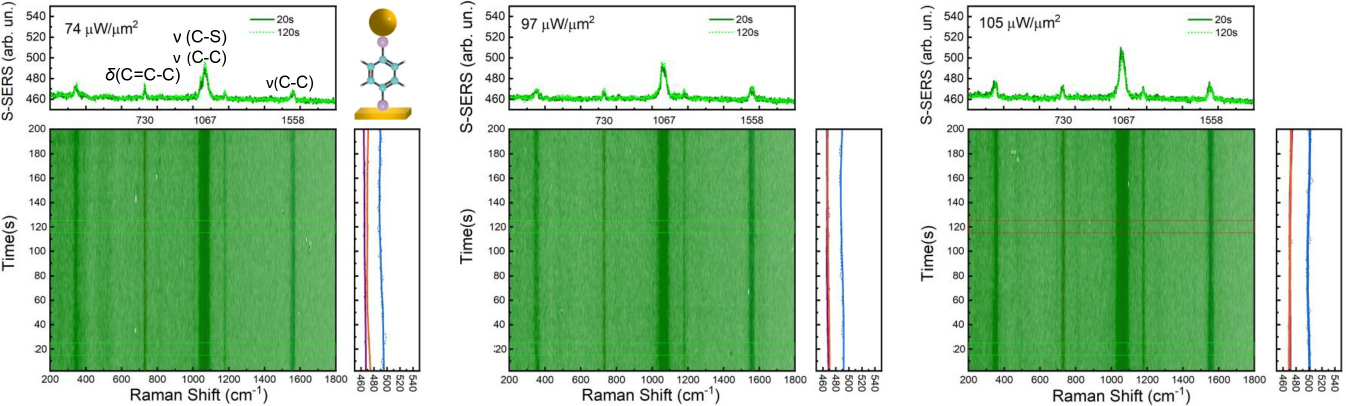}
\caption{
Temporal trace of the S-SERS spectrum measured on a BDT-sandwiched NPoM at 74, 97 and 105~$\mu\text{W}/\mu\text{m}^{2}$ excitation intensity, corresponding to the nonlinear regime, at 1.58~eV photon energy. The top panels show two cross-sections at t = 20~s (solid line) and t = 200~s (dotted line). The observed Raman peaks demonstrate outstanding temporal stability. The panels on the right show three cross-sections at the indicated spectral regions. The stable S-SERS signals (no blinking) are indicative of a lack of the geometrical displacement of Au atoms forming a pico-cavity \cite{carnegie2018room,lin2022optical} which would lead to changes of the Raman signal \cite{BenzScience2016}.
}  
\label{fig:stability}
\end{figure}
\newpage

\begin{figure}[h]%
\centering
\includegraphics[width=0.65\textwidth]{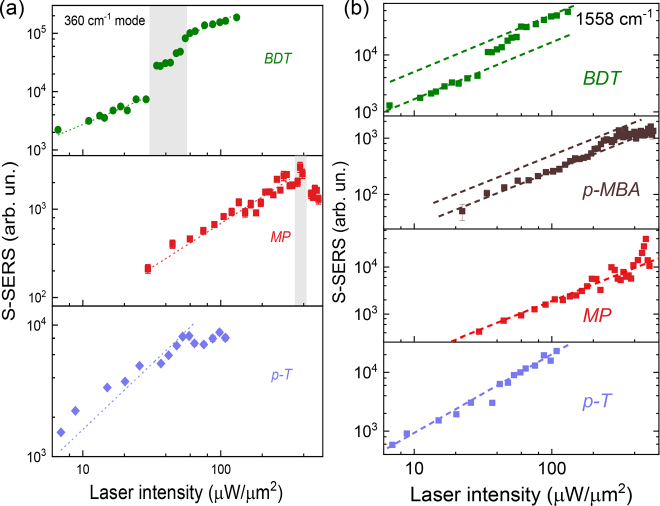}
\caption{Intensity dependence of the S-SERS signal from different molecular junctions under the resonant LSP excitation (1.58~eV, 785~nm) for vibrational modes (a) the 360~cm$^{-1}$ and (b) 1558~cm$^{-1}$. The data for the 1070~cm$^{-1}$ mode are shown in the main text. All modes demonstrate simultaneous onset of the nonlinear regime.
}
\label{fig:othermodes}
\end{figure}

\begin{figure}[h]%
\centering
\includegraphics[width=0.65\textwidth]{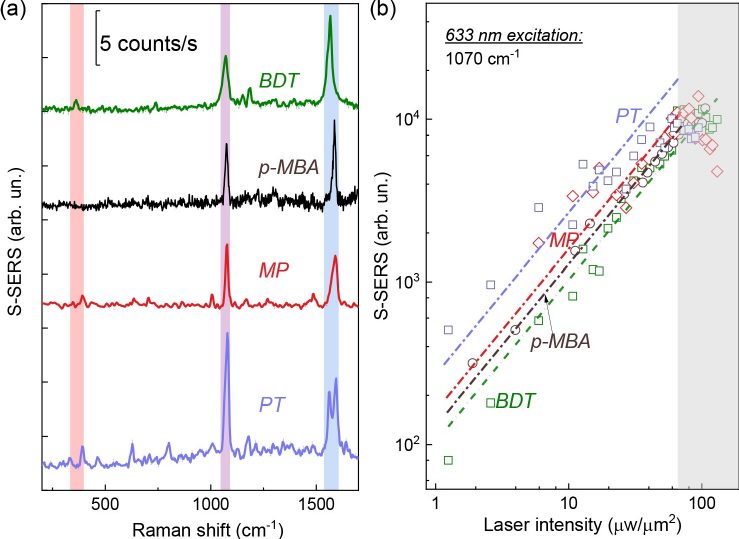}
\caption{S-SERS under off-resonant excitation (1.96~eV, 633~nm).  (a) Raman spectra of the NPoM nanojunctions with different molecules. The peaks and the shaded areas correspond to the same vibrational modes that are observed under the resonant excitation.
(b) Intensity dependence of the SERS signal for a vibrational mode of 1070~cm$^{-1}$, which demonstrates linear behaviour until the damage threshold. These results further indicate that the nonlinear Raman response is intimately related to the LSP excitation.}  
\label{fig:ramannonres}
\end{figure}
\newpage

\begin{figure}[h]%
\centering
\includegraphics[width=0.65\textwidth]{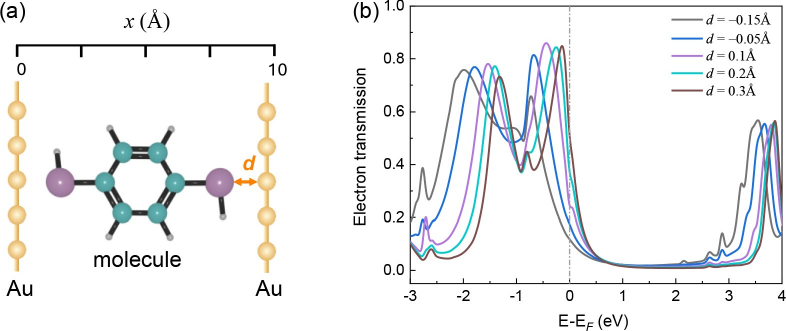}
\caption{ (a) Atomic structure of a single molecule between the two parallel planes of Au atoms (Au-molecule-Au) employed in the DFT calculations. (b) The calculated electron transmittance of BDT molecules at different distances $d$ between the Au atomic plane and the molecule. The results are compared with the experimental optical absorption spectra to identify the correct configuration of the molecule and its energy level structure (HOMO and LUMO energies).}
\label{fig:dfttrans}
\end{figure}

\begin{figure}[h]%
\centering
\includegraphics[width=0.75\textwidth]{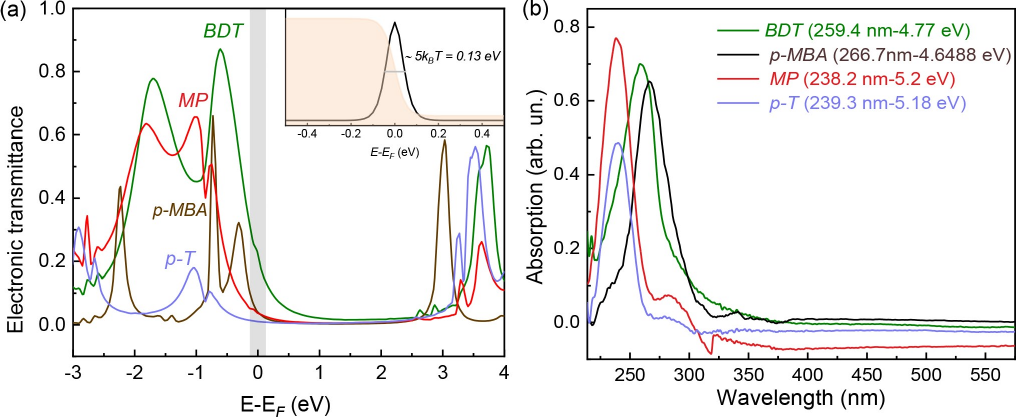}
\caption{ 
(a) The electronic transmittance of the four molecules (BDT, p-MBA, MP, and p-T) calculated using the non-equilibrium Green’s function method. The Au-BDT interatomic distance is 0.1~{\AA}. The relative conductivity of four molecules (Fig.1a) are obtained by integrating over the gray shaded region. The integration range  is chosen considering that the derivative of the Fermi-Dirac distribution $f^\prime(E-E_F)$ exhibits the most significant variations within a width of approximately \( \sim 4.53~k_B T \). To account for the majority of electronic contributions, the range was slightly extended to \( \Delta E \approx 5~k_B T \approx 0.13~eV\) at room temperature (T = 300 K). The inset illustrates the Fermi-Dirac distribution (orange shadow area) and its derivative (black line).               
(b) The absorption spectra of the molecular solutions. The wavelengths and photon energies in the legend illustrate the positions of the absorption peak. 
}  
\label{fig:dftdos}
\end{figure}

\begin{figure}[h]%
\centering
\includegraphics[width=0.7\textwidth]{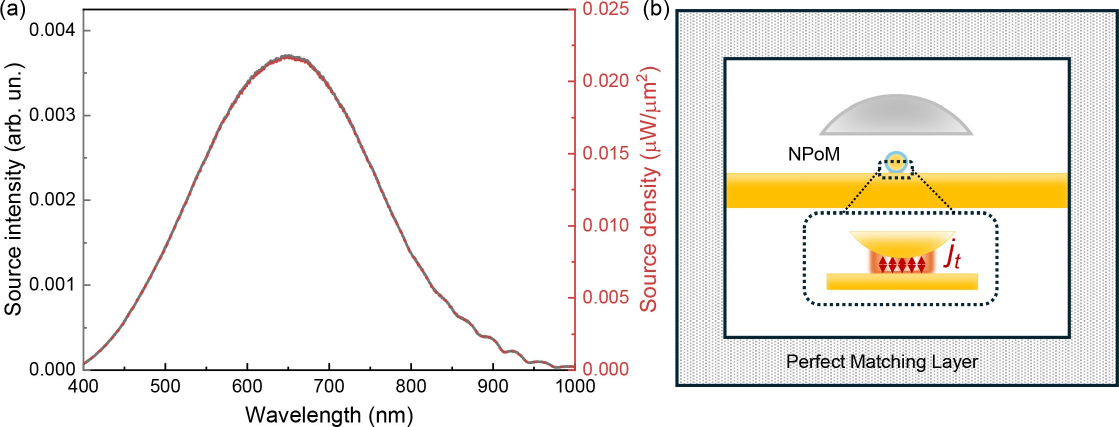}
\caption{ 
(a) Spectral density of the light source in the FDTD electromagnetic simulations using COMSOL Multiphysics. 
(b) Schematic illustration of the model geometry employed in the FDTD electromagnetic simulations.}  
\label{fig:comsolmodel}
\end{figure}

\begin{figure}[h]%
\centering
\includegraphics[width=0.55\textwidth]{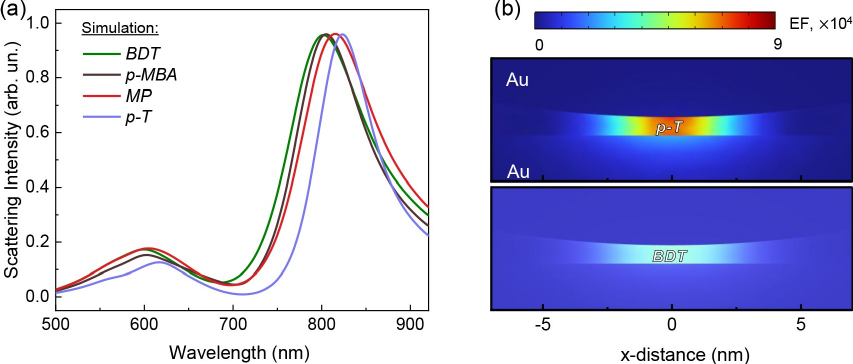}
\caption{{\bf FDTD electromagnetic simulations.}
(a) Simulated dark-field scattering spectra of NPoM plasmonic junctions with BDT, p-MBA, MP, and p-T molecules in the nanocavity. The simulation model included variations of the molecular conductance with electric field. The conductance $G_{\text{PAT}}$ obtained using the PAT formalism (conductivity $\sigma_{\text{PAT}}$) and DFT calculations (electronic transmittance) was embedded into the model. 
(b) Calculated distribution of electric field intensity $|E|^2$ (normalized to the incident laser intensity $|E_0|^2$) in the NPoM gap for the two molecular nanojunctions (BDT and p-T). The excitation wavelength is 633~nm (off-resonance). The on-resonance excitation data are presented in the main text.
}  
\label{fig:simulations}
\end{figure}

\begin{figure}[h]%
\centering
\includegraphics[width=0.45\textwidth]{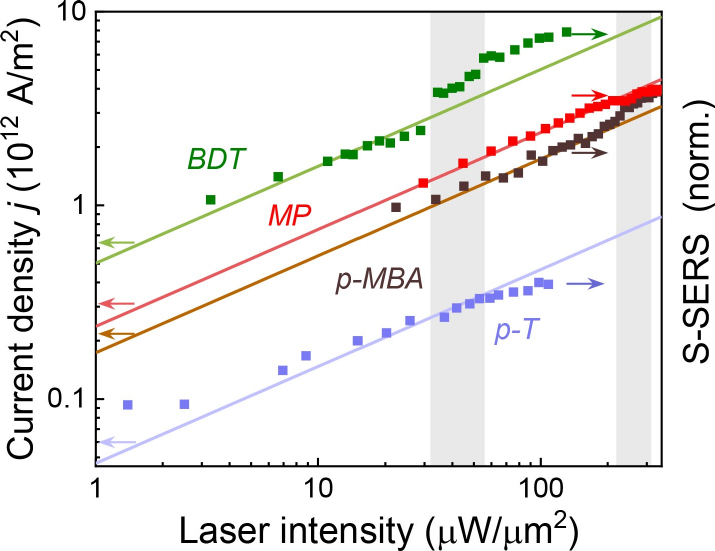}
\caption{ Calculated PAT current densities through the molecular nanojunctions as a function of the laser intensity (solid lines). The open symbols show the experimental intensity dependences of S-SERS signal measured for molecular nanojunctions. In these low-conductive nanojunctions the S-SERS power dependence exhibits a linear trend for MP and p-T molecular nanjunction. Comparing with the PAT current density, the trend of the experimental intensity dependences of the S-SERS signal matches that of the low conductivity molecular nanojunctions (MP and p-T). The PAT current densities for high-conductivity nanojunctions do not show the same increasing trend as the experimental results. The gray shaded areas highlight the nonlinear regime in the BDT and p-MBA nanojunctions response.      
}  
\label{fig: PAT current}
\end{figure}

\begin{figure}[H]%
\centering
\includegraphics[width=0.45\textwidth]{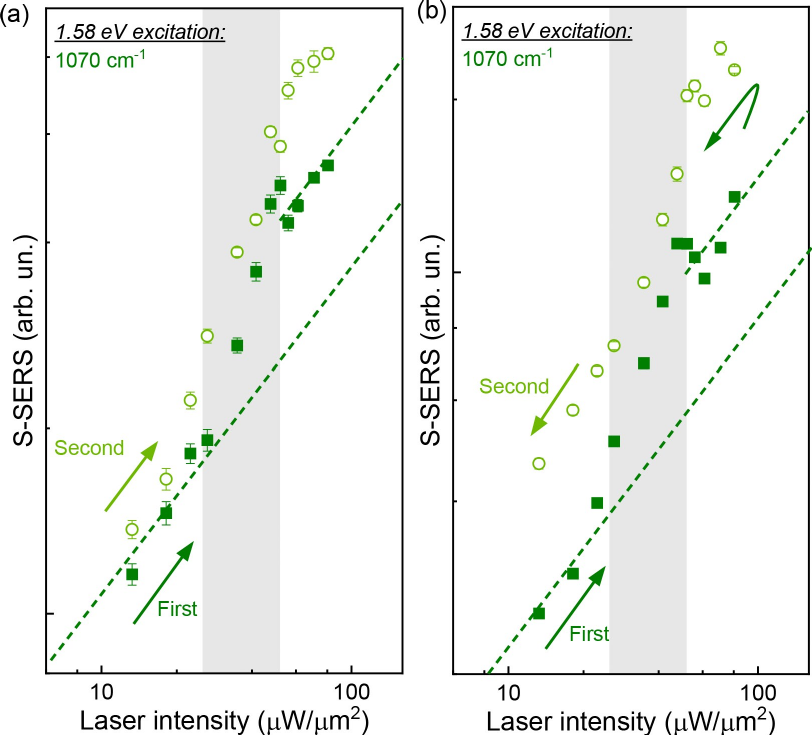}
\caption{\textbf{Reversibility test of nonlinear power dependence in BDT nanocavities.} (a) Repeated power-dependent SERS measurements conducted on the same nanoparticle, including consecutive sweeps from low to high power (15–100 $\mu$W/$\mu$m$^2$) and again from low to high power (15–100 $\mu$W/$\mu$m$^2$). (b) Power sweeps from low to high power (15–100 $\mu$W/$\mu$m$^2$) followed by high to low power (100–15 $\mu$W/$\mu$m$^2$). The second power sweep confirms the reproducibility of the S-shaped feature. Squares and circles represent data points for the first and second test, respectively, and dashed lines are linear fits. The gray shaded areas highlight the nonlinear regime in the BDT nanojunctions response.
}
\label{fig:Reversibility-785nm}
\end{figure}
\vfill
\newpage

\begin{figure}[!h]%
\centering
\includegraphics[width=0.9\textwidth]{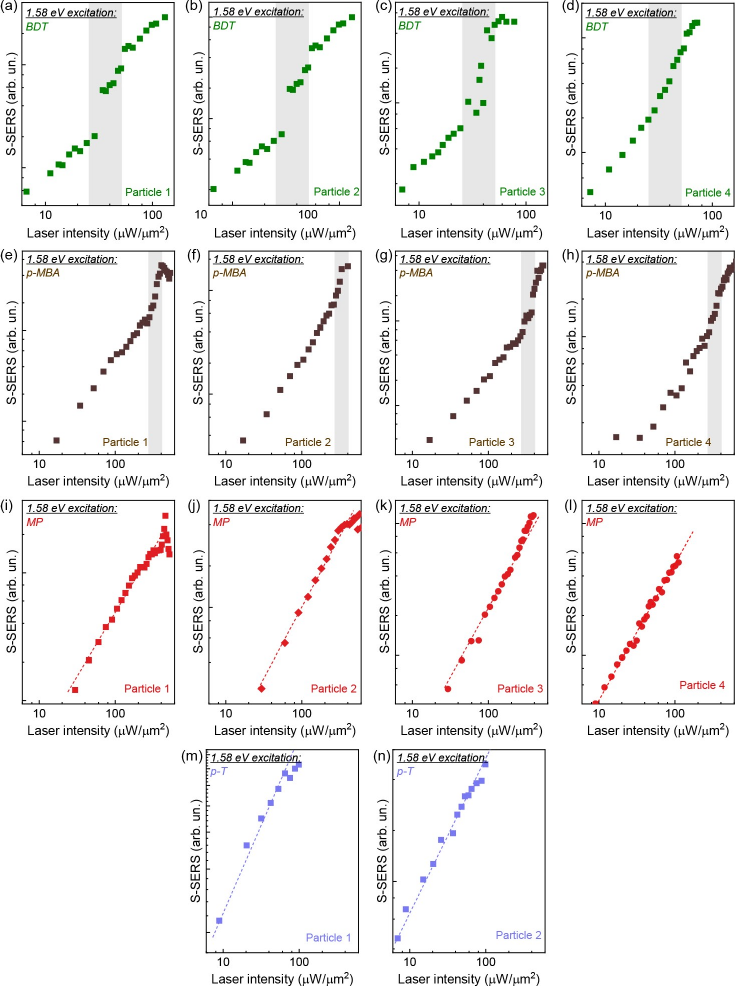}
\caption{Additional examples of the SERS power dependences from different nanojunctions (BDT, p-MBA, MP and p-T) under on-resonance (1.58 eV) excitation. The gray shaded areas highlight the nonlinear regime in the BDT nanojunctions response. Dashed lines indicate linear dependence.
} 
\label{fig:manypartuclesBDT}
\end{figure}
\newpage

\begin{figure}[h]%
\centering
\includegraphics[width=0.6\textwidth]{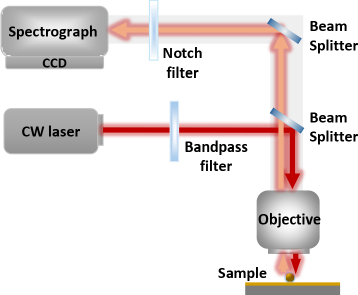}
\caption{Schematic diagram of the SERS experimental setup. The output of a continuous laser (Lead Opto-Electronics, MDL-III-581) at a wavelength of 808 nm, is focused onto the sample with a high numerical aperture microscope objective (MPlanApo 480 100$\times$, Olympus, NA $\leq$ 0.95). The Raman scattering spectra emitted by the NPoMs are collected with the same objective and then reflected into the spectrograph by two beam splitters.  A notch filter was inserted in the detection path, which removed the CW laser (Thorlabs, 808 nm $\pm$ 14 nm, FWHM $\sim$ 34 nm) and allowed the Stokes and Anti-Stokes signals to pass.
} 
\label{fig:AS_setup}
\end{figure}

\begin{figure}[h!]%
\centering
\includegraphics[width=1\textwidth]{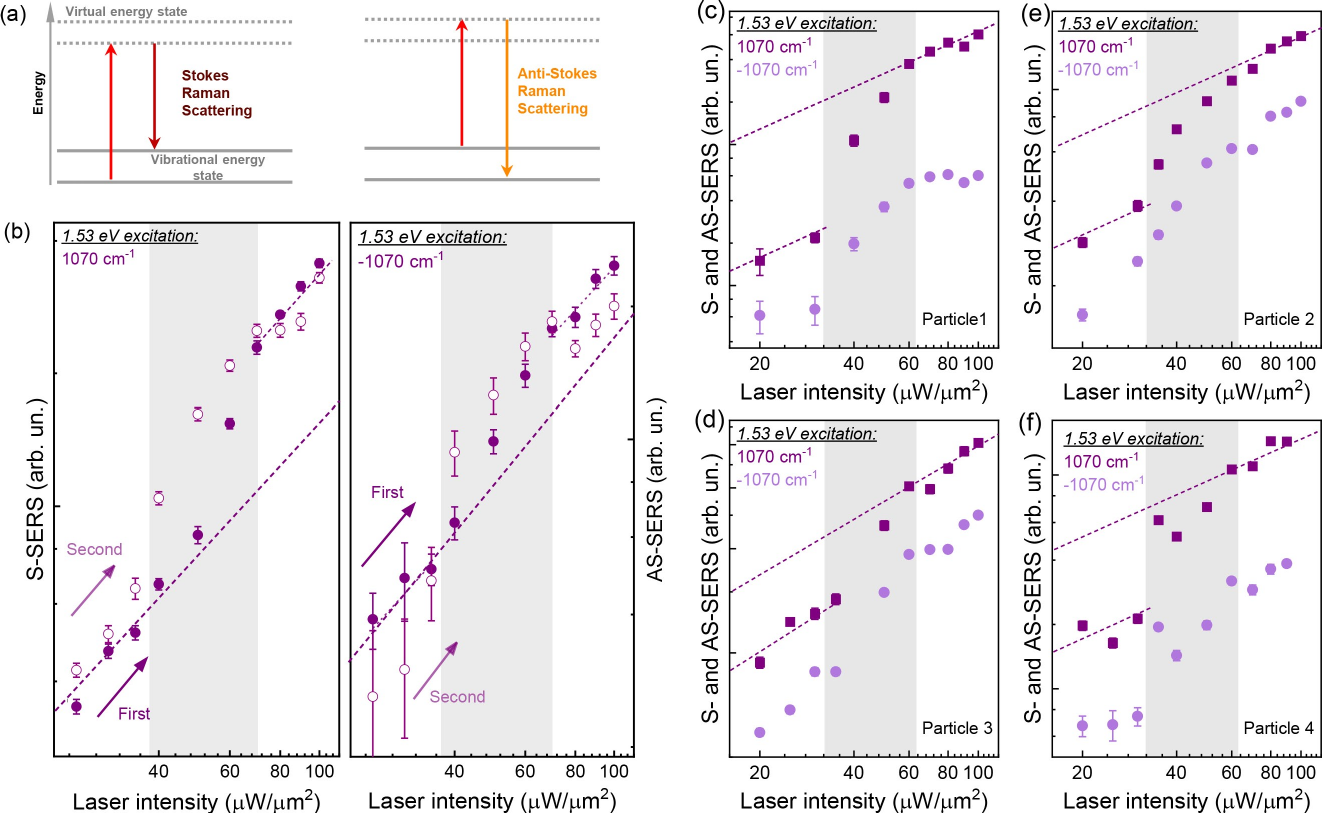}
\caption{(a) Energy diagram illustrating Stokes and Anti-Stokes Raman scattering processes with a virtual energy state and vibrational energy levels, highlighted by arrows. (b) Reversibility test of the nonlinear power dependence of BDT in the nanocavity at 1.53 eV excitation. The second power scan (from 20 to 100 $\mu$W/$\mu$m$^2$) demonstrates repeatability of the S-shaped feature. Solid and hollow dots represent the first and second power scans, respectively, with dashed lines indicating fitted curves, showing high consistency and verifying reversibility. Gray shaded areas highlight the nonlinear regime in the BDT nanojunction response. (c-f) Representative power-dependent Stokes and Anti-Stokes SERS signals from multiple BDT nanojunctions at 1.53 eV excitation (1070 cm$_1$) and 1.53 eV excitation, respectively. Dashed lines indicate linear dependence.
}
\label{fig:Reversibility}
\end{figure}
\newpage 

\begin{figure}[h]%
\centering
\includegraphics[width=0.8\textwidth]{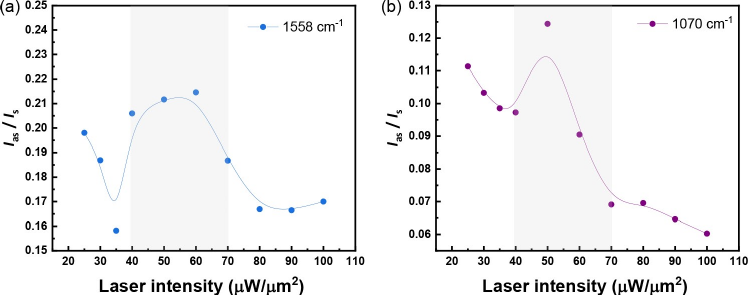}
\caption{(a-b) The intensity dependence of the ratio of the anti-Stokes to Stokes SERS signals for the 1070 cm$^{-1}$ and 1558 cm$^{-1}$ vibrational modes, showing a larger value at 50-60$\mu$W/$\mu$m$^2$, indicative of non-thermal phonon population increase. The gray shaded areas highlight the nonlinear regime in the BDT nanojunctions response.
}
\label{fig:Reversibility}
\end{figure}

\begin{figure}[h!]%
\centering
\includegraphics[width=0.9\textwidth]{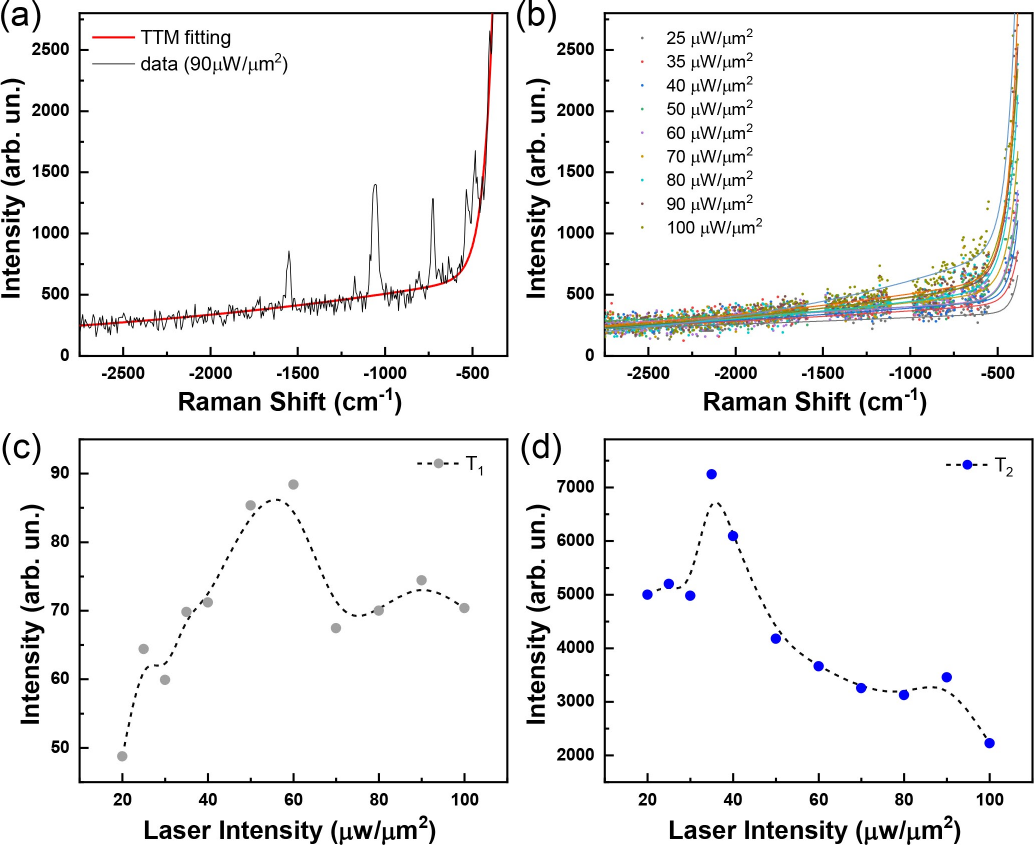}
\caption{The Two-Temperature Model (TTM) was employed to fit the background of the anti-Stokes Raman signal during optical excitation, thereby assessing the photothermal effect of the sample. The study utilized the Boltzmann distribution to describe the energy of electrons and applied a specific fitting function, incorporating two key parameters: the lattice temperature $ T_1 $ and the electron temperature $ T_2 $, to analyze the Raman signal. This function, used to fit the anti-Stokes Raman signal, is given by:$ I(\Delta\tilde{\nu}) = A_1 \exp\left(\frac{hc\Delta\tilde{\nu}}{kT_1}\right) + A_2 \exp\left(\frac{hc\Delta\tilde{\nu}}{kT_2}\right)$, where $ T_1 $ represents the lattice temperature, and $ T_2 $ represents the temperature of the electron subsystem.(a) Fitting of the anti-Stokes Raman spectral background at 90 $\mu$W/$\mu$m$^2$ using the two-temperature model (TTM). The experimental data is represented by the black line, and the fitting is represented by the red line.(b) Fitting of the anti-Stokes Raman spectral background at various laser intensities ranging from 20 to 100 $\mu$W/$\mu$m$^2$. The points are experimental data, and the line represents the fit obtained using the two-temperature model. (c) Relationship between laser intensity and lattice temperature $ T_1 $, ranging from approximately 340 to 390 K. The lattice temperature remains below 390 K within the tested power range to ensure that the molecular structure remains unchanged \cite{szczerbiński2018plasmon}. (d) Relationship between laser intensity and electron temperature $ T_2 $.
}
\label{fig:TTM}
\end{figure}

\clearpage

\clearpage

\bmhead{Acknowledgments}

Y.B., I.R., Z.G., P.T., X.L., D.J.S., and D.L. acknowledge the National Science Foundation of China through a National Excellent Young Scientists Fund grant (62022001), the Research Grants Council of Hong Kong through an ANR/RGC Joint Research Scheme grant (A-CityU101/20), the Centre for Functional Photonics of City University of Hong Kong, and the Hong Kong Branch of National Precious Metals Material Engineering Research Center (ITC Fund). A.V.Z. acknowledges support from the UK EPSRC project EP/W017075/1.

\section*{Author contributions}
Y.B. and I.R. contributed equally to this work. D.L., Y.B. and I.R. conceived the idea. D.L. designed the whole study, and supervised the project with A.V.Z.. Y.B. prepared the samples with assistance from X.L., carried out the experiments, and performed numerical simulations with assistance from P.T.. I.R. applied the Marcus electron hopping formulism to interpret the laser intensity-dependent SERS data. Z.G. performed DFT calculations under the supervision of D.J.S.. D.L. and A.V.Z. interpreted the data. Y.B. and I.R. wrote the manuscript; D.L. and A.V.Z. revised the manuscript. All other authors contributed to analyzing the data and commenting on the manuscript.

\section*{Declarations}

The authors declare no conflict of interest.

\clearpage

\input{reff.bbl}

\endgroup

\end{document}

%% file: reff.bbl

%% file: 3_Main.bbl
\begin{thebibliography}{84}
\ifx \bisbn   \undefined \def \bisbn  #1{ISBN #1}\fi
\ifx \binits  \undefined \def \binits#1{#1}\fi
\ifx \bauthor  \undefined \def \bauthor#1{#1}\fi
\ifx \batitle  \undefined \def \batitle#1{#1}\fi
\ifx \bjtitle  \undefined \def \bjtitle#1{#1}\fi
\ifx \bvolume  \undefined \def \bvolume#1{\textbf{#1}}\fi
\ifx \byear  \undefined \def \byear#1{#1}\fi
\ifx \bissue  \undefined \def \bissue#1{#1}\fi
\ifx \bfpage  \undefined \def \bfpage#1{#1}\fi
\ifx \blpage  \undefined \def \blpage #1{#1}\fi
\ifx \burl  \undefined \def \burl#1{\textsf{#1}}\fi
\ifx \doiurl  \undefined \def \doiurl#1{\url{https://doi.org/#1}}\fi
\ifx \betal  \undefined \def \betal{\textit{et al.}}\fi
\ifx \binstitute  \undefined \def \binstitute#1{#1}\fi
\ifx \binstitutionaled  \undefined \def \binstitutionaled#1{#1}\fi
\ifx \bctitle  \undefined \def \bctitle#1{#1}\fi
\ifx \beditor  \undefined \def \beditor#1{#1}\fi
\ifx \bpublisher  \undefined \def \bpublisher#1{#1}\fi
\ifx \bbtitle  \undefined \def \bbtitle#1{#1}\fi
\ifx \bedition  \undefined \def \bedition#1{#1}\fi
\ifx \bseriesno  \undefined \def \bseriesno#1{#1}\fi
\ifx \blocation  \undefined \def \blocation#1{#1}\fi
\ifx \bsertitle  \undefined \def \bsertitle#1{#1}\fi
\ifx \bsnm \undefined \def \bsnm#1{#1}\fi
\ifx \bsuffix \undefined \def \bsuffix#1{#1}\fi
\ifx \bparticle \undefined \def \bparticle#1{#1}\fi
\ifx \barticle \undefined \def \barticle#1{#1}\fi
\bibcommenthead
\ifx \bconfdate \undefined \def \bconfdate #1{#1}\fi
\ifx \botherref \undefined \def \botherref #1{#1}\fi
\ifx \url \undefined \def \url#1{\textsf{#1}}\fi
\ifx \bchapter \undefined \def \bchapter#1{#1}\fi
\ifx \bbook \undefined \def \bbook#1{#1}\fi
\ifx \bcomment \undefined \def \bcomment#1{#1}\fi
\ifx \oauthor \undefined \def \oauthor#1{#1}\fi
\ifx \citeauthoryear \undefined \def \citeauthoryear#1{#1}\fi
\ifx \endbibitem  \undefined \def \endbibitem {}\fi
\ifx \bconflocation  \undefined \def \bconflocation#1{#1}\fi
\ifx \arxivurl  \undefined \def \arxivurl#1{\textsf{#1}}\fi
\csname PreBibitemsHook\endcsname

\bibitem[\protect\citeauthoryear{Batra et~al.}{2013}]{BatraNanoLett2013}
\begin{barticle}
\bauthor{\bsnm{Batra}, \binits{A.}},
\bauthor{\bsnm{Darancet}, \binits{P.}},
\bauthor{\bsnm{Chen}, \binits{Q.}},
\bauthor{\bsnm{Meisner}, \binits{J.S.}},
\bauthor{\bsnm{Widawsky}, \binits{J.R.}},
\bauthor{\bsnm{Neaton}, \binits{J.B.}},
\bauthor{\bsnm{Nuckolls}, \binits{C.}},
\bauthor{\bsnm{Venkataraman}, \binits{L.}}:
\batitle{Tuning rectification in single-molecular diodes}.
\bjtitle{Nano Letters}
\bvolume{13}(\bissue{12}),
\bfpage{6233}--\blpage{6237}
(\byear{2013})
\end{barticle}
\endbibitem

\bibitem[\protect\citeauthoryear{Capozzi et~al.}{2015}]{CapozziNatNano2015}
\begin{barticle}
\bauthor{\bsnm{Capozzi}, \binits{B.}},
\bauthor{\bsnm{Xia}, \binits{J.}},
\bauthor{\bsnm{Adak}, \binits{O.}},
\bauthor{\bsnm{Dell}, \binits{E.J.}},
\bauthor{\bsnm{Liu}, \binits{Z.-F.}},
\bauthor{\bsnm{Taylor}, \binits{J.C.}},
\bauthor{\bsnm{Neaton}, \binits{J.B.}},
\bauthor{\bsnm{Campos}, \binits{L.M.}},
\bauthor{\bsnm{Venkataraman}, \binits{L.}}:
\batitle{Single-molecule diodes with high rectification ratios through environmental control}.
\bjtitle{Nature Nanotechnology}
\bvolume{10}(\bissue{6}),
\bfpage{522}--\blpage{527}
(\byear{2015})
\end{barticle}
\endbibitem

\bibitem[\protect\citeauthoryear{Caneva et~al.}{2018}]{CanevaNatNano2018}
\begin{barticle}
\bauthor{\bsnm{Caneva}, \binits{S.}},
\bauthor{\bsnm{Gehring}, \binits{P.}},
\bauthor{\bsnm{Garc{\'i}a-Su{\'a}rez}, \binits{V.M.}},
\bauthor{\bsnm{Garc{\'i}a-Fuente}, \binits{A.}},
\bauthor{\bsnm{Stefani}, \binits{D.}},
\bauthor{\bsnm{Olavarria-Contreras}, \binits{I.J.}},
\bauthor{\bsnm{Ferrer}, \binits{J.}},
\bauthor{\bsnm{Dekker}, \binits{C.}},
\bauthor{\bsnm{Zant}, \binits{H.S.J.}}:
\batitle{Mechanically controlled quantum interference in graphene break junctions}.
\bjtitle{Nature Nanotechnology}
\bvolume{13}(\bissue{12}),
\bfpage{1126}--\blpage{1131}
(\byear{2018})
\end{barticle}
\endbibitem

\bibitem[\protect\citeauthoryear{Li et~al.}{2018}]{LiNatNano2018}
\begin{barticle}
\bauthor{\bsnm{Li}, \binits{Y.}},
\bauthor{\bsnm{Art{\'e}s}, \binits{J.M.}},
\bauthor{\bsnm{Demir}, \binits{B.}},
\bauthor{\bsnm{Gokce}, \binits{S.}},
\bauthor{\bsnm{Mohammad}, \binits{H.M.}},
\bauthor{\bsnm{Alangari}, \binits{M.}},
\bauthor{\bsnm{Anantram}, \binits{M.P.}},
\bauthor{\bsnm{Oren}, \binits{E.E.}},
\bauthor{\bsnm{Hihath}, \binits{J.}}:
\batitle{Detection and identification of genetic material via single-molecule conductance}.
\bjtitle{Nature Nanotechnology}
\bvolume{13}(\bissue{12}),
\bfpage{1167}--\blpage{1173}
(\byear{2018})
\end{barticle}
\endbibitem

\bibitem[\protect\citeauthoryear{Chen et~al.}{2021}]{ChenScience2021}
\begin{barticle}
\bauthor{\bsnm{Chen}, \binits{W.}},
\bauthor{\bsnm{Roelli}, \binits{P.}},
\bauthor{\bsnm{Hu}, \binits{H.}},
\bauthor{\bsnm{Verlekar}, \binits{S.}},
\bauthor{\bsnm{Amirtharaj}, \binits{S.P.}},
\bauthor{\bsnm{Barreda}, \binits{A.I.}},
\bauthor{\bsnm{Kippenberg}, \binits{T.J.}},
\bauthor{\bsnm{Kovylina}, \binits{M.}},
\bauthor{\bsnm{Verhagen}, \binits{E.}},
\bauthor{\bsnm{Martínez}, \binits{A.}},
\bauthor{\bsnm{Galland}, \binits{C.}}:
\batitle{Continuous-wave frequency upconversion with a molecular optomechanical nanocavity}.
\bjtitle{Science}
\bvolume{374}(\bissue{6572}),
\bfpage{1264}--\blpage{1267}
(\byear{2021})
\end{barticle}
\endbibitem

\bibitem[\protect\citeauthoryear{Xomalis et~al.}{2021}]{XomalisScience2021}
\begin{barticle}
\bauthor{\bsnm{Xomalis}, \binits{A.}},
\bauthor{\bsnm{Zheng}, \binits{X.}},
\bauthor{\bsnm{Chikkaraddy}, \binits{R.}},
\bauthor{\bsnm{Koczor-Benda}, \binits{Z.}},
\bauthor{\bsnm{Miele}, \binits{E.}},
\bauthor{\bsnm{Rosta}, \binits{E.}},
\bauthor{\bsnm{Vandenbosch}, \binits{G.A.E.}},
\bauthor{\bsnm{Martínez}, \binits{A.}},
\bauthor{\bsnm{Baumberg}, \binits{J.J.}}:
\batitle{Detecting mid-infrared light by molecular frequency upconversion in dual-wavelength nanoantennas}.
\bjtitle{Science}
\bvolume{374}(\bissue{6572}),
\bfpage{1268}--\blpage{1271}
(\byear{2021})
\end{barticle}
\endbibitem

\bibitem[\protect\citeauthoryear{Wang et~al.}{2018}]{wang2018reactive}
\begin{barticle}
\bauthor{\bsnm{Wang}, \binits{P.}},
\bauthor{\bsnm{Krasavin}, \binits{A.V.}},
\bauthor{\bsnm{Nasir}, \binits{M.E.}},
\bauthor{\bsnm{Dickson}, \binits{W.}},
\bauthor{\bsnm{Zayats}, \binits{A.V.}}:
\batitle{Reactive tunnel junctions in electrically driven plasmonic nanorod metamaterials}.
\bjtitle{Nature nanotechnology}
\bvolume{13}(\bissue{2}),
\bfpage{159}--\blpage{164}
(\byear{2018})
\end{barticle}
\endbibitem

\bibitem[\protect\citeauthoryear{Thoss and Evers}{2018}]{ThossJCP2018}
\begin{barticle}
\bauthor{\bsnm{Thoss}, \binits{M.}},
\bauthor{\bsnm{Evers}, \binits{F.}}:
\batitle{{Perspective: Theory of quantum transport in molecular junctions}}.
\bjtitle{The Journal of Chemical Physics}
\bvolume{148}(\bissue{3}),
\bfpage{030901}
(\byear{2018})
\end{barticle}
\endbibitem

\bibitem[\protect\citeauthoryear{Zhu and Crozier}{2014}]{ZhuNatComm2014}
\begin{barticle}
\bauthor{\bsnm{Zhu}, \binits{W.}},
\bauthor{\bsnm{Crozier}, \binits{K.B.}}:
\batitle{Quantum mechanical limit to plasmonic enhancement as observed by surface-enhanced raman scattering}.
\bjtitle{Nature Communications}
\bvolume{5}(\bissue{1}),
\bfpage{5228}
(\byear{2014})
\end{barticle}
\endbibitem

\bibitem[\protect\citeauthoryear{Roelli et~al.}{2016}]{RoelliNatNano2016}
\begin{barticle}
\bauthor{\bsnm{Roelli}, \binits{P.}},
\bauthor{\bsnm{Galland}, \binits{C.}},
\bauthor{\bsnm{Piro}, \binits{N.}},
\bauthor{\bsnm{Kippenberg}, \binits{T.J.}}:
\batitle{Molecular cavity optomechanics as a theory of plasmon-enhanced raman scattering}.
\bjtitle{Nature Nanotechnology}
\bvolume{11}(\bissue{2}),
\bfpage{164}--\blpage{169}
(\byear{2016})
\end{barticle}
\endbibitem

\bibitem[\protect\citeauthoryear{Schmidt et~al.}{2016}]{SchmidtACSNano2016}
\begin{barticle}
\bauthor{\bsnm{Schmidt}, \binits{M.K.}},
\bauthor{\bsnm{Esteban}, \binits{R.}},
\bauthor{\bsnm{González-Tudela}, \binits{A.}},
\bauthor{\bsnm{Giedke}, \binits{G.}},
\bauthor{\bsnm{Aizpurua}, \binits{J.}}:
\batitle{Quantum mechanical description of raman scattering from molecules in plasmonic cavities}.
\bjtitle{ACS Nano}
\bvolume{10}(\bissue{6}),
\bfpage{6291}--\blpage{6298}
(\byear{2016})
\end{barticle}
\endbibitem

\bibitem[\protect\citeauthoryear{Cirera et~al.}{2022}]{cirera2022charge}
\begin{barticle}
\bauthor{\bsnm{Cirera}, \binits{B.}},
\bauthor{\bsnm{Litman}, \binits{Y.}},
\bauthor{\bsnm{Lin}, \binits{C.}},
\bauthor{\bsnm{Akkoush}, \binits{A.}},
\bauthor{\bsnm{Hammud}, \binits{A.}},
\bauthor{\bsnm{Wolf}, \binits{M.}},
\bauthor{\bsnm{Rossi}, \binits{M.}},
\bauthor{\bsnm{Kumagai}, \binits{T.}}:
\batitle{Charge transfer-mediated dramatic enhancement of raman scattering upon molecular point contact formation}.
\bjtitle{Nano Letters}
\bvolume{22}(\bissue{6}),
\bfpage{2170}--\blpage{2176}
(\byear{2022})
\end{barticle}
\endbibitem

\bibitem[\protect\citeauthoryear{Kippenberg et~al.}{2005}]{KippenbergPRL2005}
\begin{barticle}
\bauthor{\bsnm{Kippenberg}, \binits{T.J.}},
\bauthor{\bsnm{Rokhsari}, \binits{H.}},
\bauthor{\bsnm{Carmon}, \binits{T.}},
\bauthor{\bsnm{Scherer}, \binits{A.}},
\bauthor{\bsnm{Vahala}, \binits{K.J.}}:
\batitle{Analysis of radiation-pressure induced mechanical oscillation of an optical microcavity}.
\bjtitle{Physical Review Letters}
\bvolume{95},
\bfpage{033901}
(\byear{2005})
\end{barticle}
\endbibitem

\bibitem[\protect\citeauthoryear{Schliesser et~al.}{2006}]{SchliesserPRL2006}
\begin{barticle}
\bauthor{\bsnm{Schliesser}, \binits{A.}},
\bauthor{\bsnm{Del'Haye}, \binits{P.}},
\bauthor{\bsnm{Nooshi}, \binits{N.}},
\bauthor{\bsnm{Vahala}, \binits{K.J.}},
\bauthor{\bsnm{Kippenberg}, \binits{T.J.}}:
\batitle{Radiation pressure cooling of a micromechanical oscillator using dynamical backaction}.
\bjtitle{Physical Review Letters}
\bvolume{97},
\bfpage{243905}
(\byear{2006})
\end{barticle}
\endbibitem

\bibitem[\protect\citeauthoryear{Teufel et~al.}{2008}]{TeufelPRL2008}
\begin{barticle}
\bauthor{\bsnm{Teufel}, \binits{J.D.}},
\bauthor{\bsnm{Harlow}, \binits{J.W.}},
\bauthor{\bsnm{Regal}, \binits{C.A.}},
\bauthor{\bsnm{Lehnert}, \binits{K.W.}}:
\batitle{Dynamical backaction of microwave fields on a nanomechanical oscillator}.
\bjtitle{Physical Review Letters}
\bvolume{101},
\bfpage{197203}
(\byear{2008})
\end{barticle}
\endbibitem

\bibitem[\protect\citeauthoryear{Benz et~al.}{2016}]{BenzScience2016}
\begin{barticle}
\bauthor{\bsnm{Benz}, \binits{F.}},
\bauthor{\bsnm{Schmidt}, \binits{M.K.}},
\bauthor{\bsnm{Dreismann}, \binits{A.}},
\bauthor{\bsnm{Chikkaraddy}, \binits{R.}},
\bauthor{\bsnm{Zhang}, \binits{Y.}},
\bauthor{\bsnm{Demetriadou}, \binits{A.}},
\bauthor{\bsnm{Carnegie}, \binits{C.}},
\bauthor{\bsnm{Ohadi}, \binits{H.}},
\bauthor{\bsnm{Nijs}, \binits{B.}},
\bauthor{\bsnm{Esteban}, \binits{R.}},
\bauthor{\bsnm{Aizpurua}, \binits{J.}},
\bauthor{\bsnm{Baumberg}, \binits{J.J.}}:
\batitle{Single-molecule optomechanics in “picocavities”}.
\bjtitle{Science}
\bvolume{354}(\bissue{6313}),
\bfpage{726}--\blpage{729}
(\byear{2016})
\end{barticle}
\endbibitem

\bibitem[\protect\citeauthoryear{Lombardi et~al.}{2018}]{LombardiPRX2018}
\begin{barticle}
\bauthor{\bsnm{Lombardi}, \binits{A.}},
\bauthor{\bsnm{Schmidt}, \binits{M.K.}},
\bauthor{\bsnm{Weller}, \binits{L.}},
\bauthor{\bsnm{Deacon}, \binits{W.M.}},
\bauthor{\bsnm{Benz}, \binits{F.}},
\bauthor{\bsnm{Nijs}, \binits{B.}},
\bauthor{\bsnm{Aizpurua}, \binits{J.}},
\bauthor{\bsnm{Baumberg}, \binits{J.J.}}:
\batitle{Pulsed molecular optomechanics in plasmonic nanocavities: From nonlinear vibrational instabilities to bond-breaking}.
\bjtitle{Physical Review X}
\bvolume{8},
\bfpage{011016}
(\byear{2018})
\end{barticle}
\endbibitem

\bibitem[\protect\citeauthoryear{Jakob et~al.}{2023}]{JakobNatComm2023}
\begin{barticle}
\bauthor{\bsnm{Jakob}, \binits{L.A.}},
\bauthor{\bsnm{Deacon}, \binits{W.M.}},
\bauthor{\bsnm{Zhang}, \binits{Y.}},
\bauthor{\bsnm{Nijs}, \binits{B.}},
\bauthor{\bsnm{Pavlenko}, \binits{E.}},
\bauthor{\bsnm{Hu}, \binits{S.}},
\bauthor{\bsnm{Carnegie}, \binits{C.}},
\bauthor{\bsnm{Neuman}, \binits{T.}},
\bauthor{\bsnm{Esteban}, \binits{R.}},
\bauthor{\bsnm{Aizpurua}, \binits{J.}},
\bauthor{\bsnm{Baumberg}, \binits{J.J.}}:
\batitle{Giant optomechanical spring effect in plasmonic nano- and picocavities probed by surface-enhanced raman scattering}.
\bjtitle{Nature Communications}
\bvolume{14}(\bissue{1}),
\bfpage{3291}
(\byear{2023})
\end{barticle}
\endbibitem

\bibitem[\protect\citeauthoryear{Xu et~al.}{2022}]{xu2022phononic}
\begin{barticle}
\bauthor{\bsnm{Xu}, \binits{Y.}},
\bauthor{\bsnm{Hu}, \binits{H.}},
\bauthor{\bsnm{Chen}, \binits{W.}},
\bauthor{\bsnm{Suo}, \binits{P.}},
\bauthor{\bsnm{Zhang}, \binits{Y.}},
\bauthor{\bsnm{Zhang}, \binits{S.}},
\bauthor{\bsnm{Xu}, \binits{H.}}:
\batitle{Phononic cavity optomechanics of atomically thin crystal in plasmonic nanocavity}.
\bjtitle{ACS Nano}
\bvolume{16}(\bissue{8}),
\bfpage{12711}--\blpage{12719}
(\byear{2022})
\end{barticle}
\endbibitem

\bibitem[\protect\citeauthoryear{Hu et~al.}{2019}]{hu2019closely}
\begin{barticle}
\bauthor{\bsnm{Hu}, \binits{H.}},
\bauthor{\bsnm{Zhang}, \binits{S.}},
\bauthor{\bsnm{Xu}, \binits{H.}}:
\batitle{Closely packed metallic nanocuboid dimer allowing plasmomechanical strong coupling}.
\bjtitle{Physical Review A}
\bvolume{99}(\bissue{3}),
\bfpage{033815}
(\byear{2019})
\end{barticle}
\endbibitem

\bibitem[\protect\citeauthoryear{Kippenberg and Vahala}{2008}]{kippenberg2008cavity}
\begin{barticle}
\bauthor{\bsnm{Kippenberg}, \binits{T.J.}},
\bauthor{\bsnm{Vahala}, \binits{K.J.}}:
\batitle{Cavity optomechanics: back-action at the mesoscale}.
\bjtitle{Science}
\bvolume{321}(\bissue{5893}),
\bfpage{1172}--\blpage{1176}
(\byear{2008})
\end{barticle}
\endbibitem

\bibitem[\protect\citeauthoryear{Vahala et~al.}{2009}]{vahala2009phonon}
\begin{barticle}
\bauthor{\bsnm{Vahala}, \binits{K.}},
\bauthor{\bsnm{Herrmann}, \binits{M.}},
\bauthor{\bsnm{Kn{\"u}nz}, \binits{S.}},
\bauthor{\bsnm{Batteiger}, \binits{V.}},
\bauthor{\bsnm{Saathoff}, \binits{G.}},
\bauthor{\bsnm{H{\"a}nsch}, \binits{T.}},
\bauthor{\bsnm{Udem}, \binits{T.}}:
\batitle{A phonon laser}.
\bjtitle{Nature Physics}
\bvolume{5}(\bissue{9}),
\bfpage{682}--\blpage{686}
(\byear{2009})
\end{barticle}
\endbibitem

\bibitem[\protect\citeauthoryear{Czerniuk et~al.}{2014}]{czerniuk2014lasing}
\begin{barticle}
\bauthor{\bsnm{Czerniuk}, \binits{T.}},
\bauthor{\bsnm{Br{\"u}ggemann}, \binits{C.}},
\bauthor{\bsnm{Tepper}, \binits{J.}},
\bauthor{\bsnm{Brodbeck}, \binits{S.}},
\bauthor{\bsnm{Schneider}, \binits{C.}},
\bauthor{\bsnm{Kamp}, \binits{M.}},
\bauthor{\bsnm{H{\"o}fling}, \binits{S.}},
\bauthor{\bsnm{Glavin}, \binits{B.A.}},
\bauthor{\bsnm{Yakovlev}, \binits{D.R.}},
\bauthor{\bsnm{Akimov}, \binits{A.V.}}, \betal:
\batitle{Lasing from active optomechanical resonators}.
\bjtitle{Nature Communications}
\bvolume{5}(\bissue{1}),
\bfpage{4038}
(\byear{2014})
\end{barticle}
\endbibitem

\bibitem[\protect\citeauthoryear{Ness et~al.}{2001}]{ness2001coherent}
\begin{barticle}
\bauthor{\bsnm{Ness}, \binits{H.}},
\bauthor{\bsnm{Shevlin}, \binits{S.}},
\bauthor{\bsnm{Fisher}, \binits{A.}}:
\batitle{Coherent electron-phonon coupling and polaronlike transport in molecular wires}.
\bjtitle{Physical Review B}
\bvolume{63}(\bissue{12}),
\bfpage{125422}
(\byear{2001})
\end{barticle}
\endbibitem

\bibitem[\protect\citeauthoryear{Seldenthuis et~al.}{2008}]{seldenthuis2008vibrational}
\begin{barticle}
\bauthor{\bsnm{Seldenthuis}, \binits{J.S.}},
\bauthor{\bsnm{Van Der~Zant}, \binits{H.S.}},
\bauthor{\bsnm{Ratner}, \binits{M.A.}},
\bauthor{\bsnm{Thijssen}, \binits{J.M.}}:
\batitle{Vibrational excitations in weakly coupled single-molecule junctions: a computational analysis}.
\bjtitle{ACS Nano}
\bvolume{2}(\bissue{7}),
\bfpage{1445}--\blpage{1451}
(\byear{2008})
\end{barticle}
\endbibitem

\bibitem[\protect\citeauthoryear{Aspelmeyer et~al.}{2014}]{AspelmeyerRMP2014}
\begin{barticle}
\bauthor{\bsnm{Aspelmeyer}, \binits{M.}},
\bauthor{\bsnm{Kippenberg}, \binits{T.J.}},
\bauthor{\bsnm{Marquardt}, \binits{F.}}:
\batitle{Cavity optomechanics}.
\bjtitle{Reviews of Modern Physics}
\bvolume{86},
\bfpage{1391}--\blpage{1452}
(\byear{2014})
\end{barticle}
\endbibitem

\bibitem[\protect\citeauthoryear{Reed}{2008}]{ReedMaterials2014}
\begin{barticle}
\bauthor{\bsnm{Reed}, \binits{M.A.}}:
\batitle{Inelastic electron tunneling spectroscopy}.
\bjtitle{Materials Today}
\bvolume{11}(\bissue{11}),
\bfpage{46}--\blpage{50}
(\byear{2008})
\end{barticle}
\endbibitem

\bibitem[\protect\citeauthoryear{Eickhoff et~al.}{2020}]{EickhoffPRB2020}
\begin{barticle}
\bauthor{\bsnm{Eickhoff}, \binits{F.}},
\bauthor{\bsnm{Kolodzeiski}, \binits{E.}},
\bauthor{\bsnm{Esat}, \binits{T.}},
\bauthor{\bsnm{Fournier}, \binits{N.}},
\bauthor{\bsnm{Wagner}, \binits{C.}},
\bauthor{\bsnm{Deilmann}, \binits{T.}},
\bauthor{\bsnm{Temirov}, \binits{R.}},
\bauthor{\bsnm{Rohlfing}, \binits{M.}},
\bauthor{\bsnm{Tautz}, \binits{F.S.}},
\bauthor{\bsnm{Anders}, \binits{F.B.}}:
\batitle{Inelastic electron tunneling spectroscopy for probing strongly correlated many-body systems by scanning tunneling microscopy}.
\bjtitle{Physical Review B}
\bvolume{101},
\bfpage{125405}
(\byear{2020})
\end{barticle}
\endbibitem

\bibitem[\protect\citeauthoryear{Kneipp et~al.}{1996}]{KneippPRL1996}
\begin{barticle}
\bauthor{\bsnm{Kneipp}, \binits{K.}},
\bauthor{\bsnm{Wang}, \binits{Y.}},
\bauthor{\bsnm{Kneipp}, \binits{H.}},
\bauthor{\bsnm{Itzkan}, \binits{I.}},
\bauthor{\bsnm{Dasari}, \binits{R.R.}},
\bauthor{\bsnm{Feld}, \binits{M.S.}}:
\batitle{Population pumping of excited vibrational states by spontaneous surface-enhanced raman scattering}.
\bjtitle{Physical Review Letters}
\bvolume{76},
\bfpage{2444}--\blpage{2447}
(\byear{1996})
\end{barticle}
\endbibitem

\bibitem[\protect\citeauthoryear{Qian et~al.}{2018}]{qian2018efficient}
\begin{barticle}
\bauthor{\bsnm{Qian}, \binits{H.}},
\bauthor{\bsnm{Hsu}, \binits{S.-W.}},
\bauthor{\bsnm{Gurunatha}, \binits{K.}},
\bauthor{\bsnm{Riley}, \binits{C.T.}},
\bauthor{\bsnm{Zhao}, \binits{J.}},
\bauthor{\bsnm{Lu}, \binits{D.}},
\bauthor{\bsnm{Tao}, \binits{A.R.}},
\bauthor{\bsnm{Liu}, \binits{Z.}}:
\batitle{Efficient light generation from enhanced inelastic electron tunnelling}.
\bjtitle{Nature Photonics}
\bvolume{12}(\bissue{8}),
\bfpage{485}--\blpage{488}
(\byear{2018})
\end{barticle}
\endbibitem

\bibitem[\protect\citeauthoryear{Muniain et~al.}{2024}]{MuniainPRX2024}
\begin{barticle}
\bauthor{\bsnm{Muniain}, \binits{U.}},
\bauthor{\bsnm{Esteban}, \binits{R.}},
\bauthor{\bsnm{Aizpurua}, \binits{J.}},
\bauthor{\bsnm{Greffet}, \binits{J.-J.}}:
\batitle{Unified treatment of light emission by inelastic tunneling: Interaction of electrons and photons beyond the gap}.
\bjtitle{Physical Review X}
\bvolume{14},
\bfpage{021017}
(\byear{2024})
\end{barticle}
\endbibitem

\bibitem[\protect\citeauthoryear{Thon et~al.}{2004}]{ThonAPA2004}
\begin{barticle}
\bauthor{\bsnm{Thon}, \binits{A.}},
\bauthor{\bsnm{Merschdorf}, \binits{M.}},
\bauthor{\bsnm{Pfeiffer}, \binits{W.}},
\bauthor{\bsnm{Klamroth}, \binits{T.}},
\bauthor{\bsnm{Saalfrank}, \binits{P.}},
\bauthor{\bsnm{Diesing}, \binits{D.}}:
\batitle{Photon-assisted tunneling versus tunneling of excited electrons in metal--insulator--metal junctions}.
\bjtitle{Applied Physics A}
\bvolume{78}(\bissue{2}),
\bfpage{189}--\blpage{199}
(\byear{2004})
\end{barticle}
\endbibitem

\bibitem[\protect\citeauthoryear{Stolz et~al.}{2014}]{StolzNanoLett2014}
\begin{barticle}
\bauthor{\bsnm{Stolz}, \binits{A.}},
\bauthor{\bsnm{Berthelot}, \binits{J.}},
\bauthor{\bsnm{Mennemanteuil}, \binits{M.-M.}},
\bauthor{\bsnm{Francs}, \binits{G.}},
\bauthor{\bsnm{Markey}, \binits{L.}},
\bauthor{\bsnm{Meunier}, \binits{V.}},
\bauthor{\bsnm{Bouhelier}, \binits{A.}}:
\batitle{Nonlinear photon-assisted tunneling transport in optical gap antennas}.
\bjtitle{Nano Letters}
\bvolume{14}(\bissue{5}),
\bfpage{2330}--\blpage{2338}
(\byear{2014})
\end{barticle}
\endbibitem

\bibitem[\protect\citeauthoryear{Fung et~al.}{2017}]{FungNanoLett2017}
\begin{barticle}
\bauthor{\bsnm{Fung}, \binits{E.-D.}},
\bauthor{\bsnm{Adak}, \binits{O.}},
\bauthor{\bsnm{Lovat}, \binits{G.}},
\bauthor{\bsnm{Scarabelli}, \binits{D.}},
\bauthor{\bsnm{Venkataraman}, \binits{L.}}:
\batitle{Too hot for photon-assisted transport: Hot-electrons dominate conductance enhancement in illuminated single-molecule junctions}.
\bjtitle{Nano Letters}
\bvolume{17}(\bissue{2}),
\bfpage{1255}--\blpage{1261}
(\byear{2017})
\end{barticle}
\endbibitem

\bibitem[\protect\citeauthoryear{Kos et~al.}{2021}]{kos2021quantum}
\begin{barticle}
\bauthor{\bsnm{Kos}, \binits{D.}},
\bauthor{\bsnm{Assumpcao}, \binits{D.R.}},
\bauthor{\bsnm{Guo}, \binits{C.}},
\bauthor{\bsnm{Baumberg}, \binits{J.J.}}:
\batitle{Quantum tunneling induced optical rectification and plasmon-enhanced photocurrent in nanocavity molecular junctions}.
\bjtitle{ACS Nano}
\bvolume{15}(\bissue{9}),
\bfpage{14535}--\blpage{14543}
(\byear{2021})
\end{barticle}
\endbibitem

\bibitem[\protect\citeauthoryear{Cui et~al.}{2018}]{cui2018molecular}
\begin{barticle}
\bauthor{\bsnm{Cui}, \binits{X.}},
\bauthor{\bsnm{Qin}, \binits{F.}},
\bauthor{\bsnm{Lai}, \binits{Y.}},
\bauthor{\bsnm{Wang}, \binits{H.}},
\bauthor{\bsnm{Shao}, \binits{L.}},
\bauthor{\bsnm{Chen}, \binits{H.}},
\bauthor{\bsnm{Wang}, \binits{J.}},
\bauthor{\bsnm{Lin}, \binits{H.-q.}}:
\batitle{Molecular tunnel junction-controlled high-order charge transfer plasmon and fano resonances}.
\bjtitle{ACS Nano}
\bvolume{12}(\bissue{12}),
\bfpage{12541}--\blpage{12550}
(\byear{2018})
\end{barticle}
\endbibitem

\bibitem[\protect\citeauthoryear{Suzuki et~al.}{2016}]{suzuki2016effect}
\begin{barticle}
\bauthor{\bsnm{Suzuki}, \binits{S.}},
\bauthor{\bsnm{Kaneko}, \binits{S.}},
\bauthor{\bsnm{Fujii}, \binits{S.}},
\bauthor{\bsnm{Marqu{\'e}s-Gonz{\'a}lez}, \binits{S.}},
\bauthor{\bsnm{Nishino}, \binits{T.}},
\bauthor{\bsnm{Kiguchi}, \binits{M.}}:
\batitle{Effect of the molecule--metal interface on the surface-enhanced raman scattering of 1, 4-benzenedithiol}.
\bjtitle{The Journal of Physical Chemistry C}
\bvolume{120}(\bissue{2}),
\bfpage{1038}--\blpage{1042}
(\byear{2016})
\end{barticle}
\endbibitem

\bibitem[\protect\citeauthoryear{Caligiuri et~al.}{2020}]{caligiuri2020biodegradable}
\begin{barticle}
\bauthor{\bsnm{Caligiuri}, \binits{V.}},
\bauthor{\bsnm{Tedeschi}, \binits{G.}},
\bauthor{\bsnm{Palei}, \binits{M.}},
\bauthor{\bsnm{Miscuglio}, \binits{M.}},
\bauthor{\bsnm{Martin-Garcia}, \binits{B.}},
\bauthor{\bsnm{Guzman-Puyol}, \binits{S.}},
\bauthor{\bsnm{Hedayati}, \binits{M.K.}},
\bauthor{\bsnm{Kristensen}, \binits{A.}},
\bauthor{\bsnm{Athanassiou}, \binits{A.}},
\bauthor{\bsnm{Cingolani}, \binits{R.}}, \betal:
\batitle{Biodegradable and insoluble cellulose photonic crystals and metasurfaces}.
\bjtitle{ACS Nano}
\bvolume{14}(\bissue{8}),
\bfpage{9502}--\blpage{9511}
(\byear{2020})
\end{barticle}
\endbibitem

\bibitem[\protect\citeauthoryear{Aravind and Metiu}{1982}]{AravindJPC1982}
\begin{barticle}
\bauthor{\bsnm{Aravind}, \binits{P.K.}},
\bauthor{\bsnm{Metiu}, \binits{H.}}:
\batitle{Use of a perfectly conducting sphere to excite the plasmon of a flat surface. 1. calculation of the local field with applications to surface-enhanced spectroscopy}.
\bjtitle{The Journal of Physical Chemistry}
\bvolume{86}(\bissue{26}),
\bfpage{5076}--\blpage{5084}
(\byear{1982})
\end{barticle}
\endbibitem

\bibitem[\protect\citeauthoryear{Baumberg et~al.}{2019}]{BaumbergNatMater2019}
\begin{barticle}
\bauthor{\bsnm{Baumberg}, \binits{J.J.}},
\bauthor{\bsnm{Aizpurua}, \binits{J.}},
\bauthor{\bsnm{Mikkelsen}, \binits{M.H.}},
\bauthor{\bsnm{Smith}, \binits{D.R.}}:
\batitle{Extreme nanophotonics from ultrathin metallic gaps}.
\bjtitle{Nature Materials}
\bvolume{18}(\bissue{7}),
\bfpage{668}--\blpage{678}
(\byear{2019})
\end{barticle}
\endbibitem

\bibitem[\protect\citeauthoryear{Hu et~al.}{2022}]{HuNanop2022}
\begin{barticle}
\bauthor{\bsnm{Hu}, \binits{H.}},
\bauthor{\bsnm{Xu}, \binits{Y.}},
\bauthor{\bsnm{Hu}, \binits{Z.}},
\bauthor{\bsnm{Kang}, \binits{B.}},
\bauthor{\bsnm{Zhang}, \binits{Z.}},
\bauthor{\bsnm{Sun}, \binits{J.}},
\bauthor{\bsnm{Li}, \binits{Y.}},
\bauthor{\bsnm{Xu}, \binits{H.}}:
\batitle{Nanoparticle-on-mirror pairs: building blocks for remote spectroscopies}.
\bjtitle{Nanophotonics}
\bvolume{11}(\bissue{22}),
\bfpage{5153}--\blpage{5163}
(\byear{2022})
\end{barticle}
\endbibitem

\bibitem[\protect\citeauthoryear{Thomas et~al.}{2018}]{thomas2018acid}
\begin{barticle}
\bauthor{\bsnm{Thomas}, \binits{J.C.}},
\bauthor{\bsnm{Goronzy}, \binits{D.P.}},
\bauthor{\bsnm{Serino}, \binits{A.C.}},
\bauthor{\bsnm{Auluck}, \binits{H.S.}},
\bauthor{\bsnm{Irving}, \binits{O.R.}},
\bauthor{\bsnm{Jimenez-Izal}, \binits{E.}},
\bauthor{\bsnm{Deirmenjian}, \binits{J.M.}},
\bauthor{\bsnm{Machacek}, \binits{J.}},
\bauthor{\bsnm{Sautet}, \binits{P.}},
\bauthor{\bsnm{Alexandrova}, \binits{A.N.}}, \betal:
\batitle{Acid--base control of valency within carboranedithiol self-assembled monolayers: molecules do the can-can}.
\bjtitle{ACS Nano}
\bvolume{12}(\bissue{3}),
\bfpage{2211}--\blpage{2221}
(\byear{2018})
\end{barticle}
\endbibitem

\bibitem[\protect\citeauthoryear{Benz et~al.}{2015}]{benz2015nanooptics}
\begin{barticle}
\bauthor{\bsnm{Benz}, \binits{F.}},
\bauthor{\bsnm{Tserkezis}, \binits{C.}},
\bauthor{\bsnm{Herrmann}, \binits{L.O.}},
\bauthor{\bsnm{De~Nijs}, \binits{B.}},
\bauthor{\bsnm{Sanders}, \binits{A.}},
\bauthor{\bsnm{Sigle}, \binits{D.O.}},
\bauthor{\bsnm{Pukenas}, \binits{L.}},
\bauthor{\bsnm{Evans}, \binits{S.D.}},
\bauthor{\bsnm{Aizpurua}, \binits{J.}},
\bauthor{\bsnm{Baumberg}, \binits{J.J.}}:
\batitle{Nanooptics of molecular-shunted plasmonic nanojunctions}.
\bjtitle{Nano Letters}
\bvolume{15}(\bissue{1}),
\bfpage{669}--\blpage{674}
(\byear{2015})
\end{barticle}
\endbibitem

\bibitem[\protect\citeauthoryear{Park and Kim}{2010}]{park2010charge}
\begin{barticle}
\bauthor{\bsnm{Park}, \binits{W.-H.}},
\bauthor{\bsnm{Kim}, \binits{Z.H.}}:
\batitle{Charge transfer enhancement in the sers of a single molecule}.
\bjtitle{Nano letters}
\bvolume{10}(\bissue{10}),
\bfpage{4040}--\blpage{4048}
(\byear{2010})
\end{barticle}
\endbibitem

\bibitem[\protect\citeauthoryear{Benz et~al.}{2015}]{benz2015generalized}
\begin{barticle}
\bauthor{\bsnm{Benz}, \binits{F.}},
\bauthor{\bsnm{Nijs}, \binits{B.}},
\bauthor{\bsnm{Tserkezis}, \binits{C.}},
\bauthor{\bsnm{Chikkaraddy}, \binits{R.}},
\bauthor{\bsnm{Sigle}, \binits{D.O.}},
\bauthor{\bsnm{Pukenas}, \binits{L.}},
\bauthor{\bsnm{Evans}, \binits{S.D.}},
\bauthor{\bsnm{Aizpurua}, \binits{J.}},
\bauthor{\bsnm{Baumberg}, \binits{J.J.}}:
\batitle{Generalized circuit model for coupled plasmonic systems}.
\bjtitle{Optics Express}
\bvolume{23}(\bissue{26}),
\bfpage{33255}--\blpage{33269}
(\byear{2015})
\end{barticle}
\endbibitem

\bibitem[\protect\citeauthoryear{Carnegie et~al.}{2018}]{carnegie2018room}
\begin{barticle}
\bauthor{\bsnm{Carnegie}, \binits{C.}},
\bauthor{\bsnm{Griffiths}, \binits{J.}},
\bauthor{\bsnm{Nijs}, \binits{B.}},
\bauthor{\bsnm{Readman}, \binits{C.}},
\bauthor{\bsnm{Chikkaraddy}, \binits{R.}},
\bauthor{\bsnm{Deacon}, \binits{W.M.}},
\bauthor{\bsnm{Zhang}, \binits{Y.}},
\bauthor{\bsnm{Szab{\'o}}, \binits{I.}},
\bauthor{\bsnm{Rosta}, \binits{E.}},
\bauthor{\bsnm{Aizpurua}, \binits{J.}}, \betal:
\batitle{Room-temperature optical picocavities below 1 nm$^3$ accessing single-atom geometries}.
\bjtitle{The Journal of Physical Chemistry Letters}
\bvolume{9}(\bissue{24}),
\bfpage{7146}--\blpage{7151}
(\byear{2018})
\end{barticle}
\endbibitem

\bibitem[\protect\citeauthoryear{Lin et~al.}{2022}]{lin2022optical}
\begin{barticle}
\bauthor{\bsnm{Lin}, \binits{Q.}},
\bauthor{\bsnm{Hu}, \binits{S.}},
\bauthor{\bsnm{F{\"o}ldes}, \binits{T.}},
\bauthor{\bsnm{Huang}, \binits{J.}},
\bauthor{\bsnm{Wright}, \binits{D.}},
\bauthor{\bsnm{Griffiths}, \binits{J.}},
\bauthor{\bsnm{Elliott}, \binits{E.}},
\bauthor{\bsnm{Nijs}, \binits{B.}},
\bauthor{\bsnm{Rosta}, \binits{E.}},
\bauthor{\bsnm{Baumberg}, \binits{J.J.}}:
\batitle{Optical suppression of energy barriers in single molecule-metal binding}.
\bjtitle{Science Advances}
\bvolume{8}(\bissue{25}),
\bfpage{9285}
(\byear{2022})
\end{barticle}
\endbibitem

\bibitem[\protect\citeauthoryear{Seldenthuis et~al.}{2008}]{seldenthuisACSNano2008}
\begin{barticle}
\bauthor{\bsnm{Seldenthuis}, \binits{J.S.}},
\bauthor{\bsnm{Zant}, \binits{H.S.J.}},
\bauthor{\bsnm{Ratner}, \binits{M.A.}},
\bauthor{\bsnm{Thijssen}, \binits{J.M.}}:
\batitle{Vibrational excitations in weakly coupled single-molecule junctions: A computational analysis}.
\bjtitle{ACS Nano}
\bvolume{2}(\bissue{7}),
\bfpage{1445}--\blpage{1451}
(\byear{2008})
\end{barticle}
\endbibitem

\bibitem[\protect\citeauthoryear{Soler et~al.}{2002}]{soler2002siesta}
\begin{barticle}
\bauthor{\bsnm{Soler}, \binits{J.M.}},
\bauthor{\bsnm{Artacho}, \binits{E.}},
\bauthor{\bsnm{Gale}, \binits{J.D.}},
\bauthor{\bsnm{Garc{\'\i}a}, \binits{A.}},
\bauthor{\bsnm{Junquera}, \binits{J.}},
\bauthor{\bsnm{Ordej{\'o}n}, \binits{P.}},
\bauthor{\bsnm{S{\'a}nchez-Portal}, \binits{D.}}:
\batitle{The siesta method for ab initio order-n materials simulation}.
\bjtitle{Journal of Physics: Condensed Matter}
\bvolume{14}(\bissue{11}),
\bfpage{2745}
(\byear{2002})
\end{barticle}
\endbibitem

\bibitem[\protect\citeauthoryear{Brandbyge et~al.}{2002}]{brandbyge2002density}
\begin{barticle}
\bauthor{\bsnm{Brandbyge}, \binits{M.}},
\bauthor{\bsnm{Mozos}, \binits{J.-L.}},
\bauthor{\bsnm{Ordej{\'o}n}, \binits{P.}},
\bauthor{\bsnm{Taylor}, \binits{J.}},
\bauthor{\bsnm{Stokbro}, \binits{K.}}:
\batitle{Density-functional method for nonequilibrium electron transport}.
\bjtitle{Physical Review B}
\bvolume{65}(\bissue{16}),
\bfpage{165401}
(\byear{2002})
\end{barticle}
\endbibitem

\bibitem[\protect\citeauthoryear{Kresse and Furthm{\"u}ller}{1996}]{kresse1996efficient}
\begin{barticle}
\bauthor{\bsnm{Kresse}, \binits{G.}},
\bauthor{\bsnm{Furthm{\"u}ller}, \binits{J.}}:
\batitle{Efficient iterative schemes for ab initio total-energy calculations using a plane-wave basis set}.
\bjtitle{Physical Review B}
\bvolume{54}(\bissue{16}),
\bfpage{11169}
(\byear{1996})
\end{barticle}
\endbibitem

\bibitem[\protect\citeauthoryear{Perdew et~al.}{1996}]{perdew1996generalized}
\begin{barticle}
\bauthor{\bsnm{Perdew}, \binits{J.P.}},
\bauthor{\bsnm{Burke}, \binits{K.}},
\bauthor{\bsnm{Ernzerhof}, \binits{M.}}:
\batitle{Generalized gradient approximation made simple}.
\bjtitle{Physical Review Letters}
\bvolume{77}(\bissue{18}),
\bfpage{3865}
(\byear{1996})
\end{barticle}
\endbibitem

\bibitem[\protect\citeauthoryear{Bl{\"o}chl}{1994}]{blochl1994projector}
\begin{barticle}
\bauthor{\bsnm{Bl{\"o}chl}, \binits{P.E.}}:
\batitle{Projector augmented-wave method}.
\bjtitle{Physical Review B}
\bvolume{50}(\bissue{24}),
\bfpage{17953}
(\byear{1994})
\end{barticle}
\endbibitem

\bibitem[\protect\citeauthoryear{Tien and Gordon}{1963}]{TienGordonPR1963}
\begin{barticle}
\bauthor{\bsnm{Tien}, \binits{P.K.}},
\bauthor{\bsnm{Gordon}, \binits{J.P.}}:
\batitle{Multiphoton process observed in the interaction of microwave fields with the tunneling between superconductor films}.
\bjtitle{Physical Review}
\bvolume{129},
\bfpage{647}--\blpage{651}
(\byear{1963})
\end{barticle}
\endbibitem

\bibitem[\protect\citeauthoryear{Galperin et~al.}{2007}]{GalperinJPCM2007}
\begin{barticle}
\bauthor{\bsnm{Galperin}, \binits{M.}},
\bauthor{\bsnm{Ratner}, \binits{M.A.}},
\bauthor{\bsnm{Nitzan}, \binits{A.}}:
\batitle{Molecular transport junctions: vibrational effects}.
\bjtitle{Journal of Physics: Condensed Matter}
\bvolume{19}(\bissue{10}),
\bfpage{103201}
(\byear{2007})
\end{barticle}
\endbibitem

\bibitem[\protect\citeauthoryear{van~der Molen and Liljeroth}{2010}]{vanderMolenJPCM2010}
\begin{barticle}
\bauthor{\bsnm{Molen}, \binits{S.J.}},
\bauthor{\bsnm{Liljeroth}, \binits{P.}}:
\batitle{Charge transport through molecular switches}.
\bjtitle{Journal of Physics: Condensed Matter}
\bvolume{22}(\bissue{13}),
\bfpage{133001}
(\byear{2010})
\end{barticle}
\endbibitem

\bibitem[\protect\citeauthoryear{Jaklevic and Lambe}{1966}]{JaklevicPRL1966}
\begin{barticle}
\bauthor{\bsnm{Jaklevic}, \binits{R.C.}},
\bauthor{\bsnm{Lambe}, \binits{J.}}:
\batitle{Molecular vibration spectra by electron tunneling}.
\bjtitle{Physical Review Letters}
\bvolume{17},
\bfpage{1139}--\blpage{1140}
(\byear{1966})
\end{barticle}
\endbibitem

\bibitem[\protect\citeauthoryear{B\"uttiker and Landauer}{1982}]{ButtikerPRL1982}
\begin{barticle}
\bauthor{\bsnm{B\"uttiker}, \binits{M.}},
\bauthor{\bsnm{Landauer}, \binits{R.}}:
\batitle{Traversal time for tunneling}.
\bjtitle{Physical Review Letters}
\bvolume{49},
\bfpage{1739}--\blpage{1742}
(\byear{1982})
\end{barticle}
\endbibitem

\bibitem[\protect\citeauthoryear{Halbritter}{1982}]{HalbritterSurf1982}
\begin{barticle}
\bauthor{\bsnm{Halbritter}, \binits{J.}}:
\batitle{On resonant tunneling}.
\bjtitle{Surface Science}
\bvolume{122}(\bissue{1}),
\bfpage{80}--\blpage{98}
(\byear{1982})
\end{barticle}
\endbibitem

\bibitem[\protect\citeauthoryear{Bending and Beasley}{1985}]{BendingPRL1985}
\begin{barticle}
\bauthor{\bsnm{Bending}, \binits{S.J.}},
\bauthor{\bsnm{Beasley}, \binits{M.R.}}:
\batitle{Transport processes via localized states in thin a-si tunnel barriers}.
\bjtitle{Physical Review Letters}
\bvolume{55},
\bfpage{324}--\blpage{327}
(\byear{1985})
\end{barticle}
\endbibitem

\bibitem[\protect\citeauthoryear{Ozaki et~al.}{1998}]{OzakiJAP1998}
\begin{barticle}
\bauthor{\bsnm{Ozaki}, \binits{S.}},
\bauthor{\bsnm{Feng}, \binits{J.M.}},
\bauthor{\bsnm{Park}, \binits{J.H.}},
\bauthor{\bsnm{Osako}, \binits{S.-i.}},
\bauthor{\bsnm{Kubo}, \binits{H.}},
\bauthor{\bsnm{Morifuji}, \binits{M.}},
\bauthor{\bsnm{Mori}, \binits{N.}},
\bauthor{\bsnm{Hamaguchi}, \binits{C.}}:
\batitle{{Observation of resonant optical–phonon assisted tunneling in asymmetric double quantum wells}}.
\bjtitle{Journal of Applied Physics}
\bvolume{83}(\bissue{2}),
\bfpage{962}--\blpage{965}
(\byear{1998})
\end{barticle}
\endbibitem

\bibitem[\protect\citeauthoryear{Yu et~al.}{2004}]{YuPRL2004}
\begin{barticle}
\bauthor{\bsnm{Yu}, \binits{L.H.}},
\bauthor{\bsnm{Keane}, \binits{Z.K.}},
\bauthor{\bsnm{Ciszek}, \binits{J.W.}},
\bauthor{\bsnm{Cheng}, \binits{L.}},
\bauthor{\bsnm{Stewart}, \binits{M.P.}},
\bauthor{\bsnm{Tour}, \binits{J.M.}},
\bauthor{\bsnm{Natelson}, \binits{D.}}:
\batitle{Inelastic electron tunneling via molecular vibrations in single-molecule transistors}.
\bjtitle{Physical Review Letters}
\bvolume{93},
\bfpage{266802}
(\byear{2004})
\end{barticle}
\endbibitem

\bibitem[\protect\citeauthoryear{Galperin et~al.}{2006}]{GalperinPRL2006}
\begin{barticle}
\bauthor{\bsnm{Galperin}, \binits{M.}},
\bauthor{\bsnm{Nitzan}, \binits{A.}},
\bauthor{\bsnm{Ratner}, \binits{M.A.}}:
\batitle{Molecular transport junctions: Current from electronic excitations in the leads}.
\bjtitle{Physical Review Letters}
\bvolume{96},
\bfpage{166803}
(\byear{2006})
\end{barticle}
\endbibitem

\bibitem[\protect\citeauthoryear{Leijnse and Wegewijs}{2008}]{LeijnsePRB2008}
\begin{barticle}
\bauthor{\bsnm{Leijnse}, \binits{M.}},
\bauthor{\bsnm{Wegewijs}, \binits{M.R.}}:
\batitle{Kinetic equations for transport through single-molecule transistors}.
\bjtitle{Physical Review B}
\bvolume{78},
\bfpage{235424}
(\byear{2008})
\end{barticle}
\endbibitem

\bibitem[\protect\citeauthoryear{Scalapino and Marcus}{1967}]{ScalapinoPRL1967}
\begin{barticle}
\bauthor{\bsnm{Scalapino}, \binits{D.J.}},
\bauthor{\bsnm{Marcus}, \binits{S.M.}}:
\batitle{Theory of inelastic electron-molecule interactions in tunnel junctions}.
\bjtitle{Physical Review Letters}
\bvolume{18},
\bfpage{459}--\blpage{461}
(\byear{1967})
\end{barticle}
\endbibitem

\bibitem[\protect\citeauthoryear{Sowa et~al.}{2018}]{SowaJPC2018}
\begin{barticle}
\bauthor{\bsnm{Sowa}, \binits{J.K.}},
\bauthor{\bsnm{Mol}, \binits{J.A.}},
\bauthor{\bsnm{Briggs}, \binits{G.A.D.}},
\bauthor{\bsnm{Gauger}, \binits{E.M.}}:
\batitle{{Beyond Marcus theory and the Landauer-Büttiker approach in molecular junctions: A unified framework}}.
\bjtitle{The Journal of Chemical Physics}
\bvolume{149}(\bissue{15}),
\bfpage{154112}
(\byear{2018})
\end{barticle}
\endbibitem

\bibitem[\protect\citeauthoryear{{Lang} and {Firsov}}{1963}]{LangFirsov1963}
\begin{barticle}
\bauthor{\bsnm{{Lang}}, \binits{I.G.}},
\bauthor{\bsnm{{Firsov}}, \binits{Y.A.}}:
\batitle{{Kinetic Theory of Semiconductors with Low Mobility}}.
\bjtitle{Soviet Journal of Experimental and Theoretical Physics}
\bvolume{16},
\bfpage{1301}
(\byear{1963})
\end{barticle}
\endbibitem

\bibitem[\protect\citeauthoryear{Maher et~al.}{2006}]{MaherJPCB2006}
\begin{barticle}
\bauthor{\bsnm{Maher}, \binits{R.C.}},
\bauthor{\bsnm{Etchegoin}, \binits{P.G.}},
\bauthor{\bsnm{Le~Ru}, \binits{E.C.}},
\bauthor{\bsnm{Cohen}, \binits{L.F.}}:
\batitle{A conclusive demonstration of vibrational pumping under surface enhanced raman scattering conditions}.
\bjtitle{The Journal of Physical Chemistry B}
\bvolume{110}(\bissue{24}),
\bfpage{11757}--\blpage{11760}
(\byear{2006})
\end{barticle}
\endbibitem

\bibitem[\protect\citeauthoryear{{Ruhman, S.} et~al.}{1987}]{RuhmanRPA1987}
\begin{barticle}
\bauthor{\bsnm{{Ruhman, S.}}},
\bauthor{\bsnm{{Joly, A.G.}}},
\bauthor{\bsnm{{Kohler, B.}}},
\bauthor{\bsnm{{Williams, L.R.}}},
\bauthor{\bsnm{{Nelson, K.A.}}}:
\batitle{Intramolecular and intermolecular dynamics in molecular liquids through femtosecond time-resolved impulsive stimulated scattering}.
\bjtitle{Revue de Physique Appliquée (Paris)}
\bvolume{22}(\bissue{12}),
\bfpage{1717}--\blpage{1734}
(\byear{1987})
\end{barticle}
\endbibitem

\bibitem[\protect\citeauthoryear{Cirera et~al.}{2022}]{CireraACSNano2022}
\begin{barticle}
\bauthor{\bsnm{Cirera}, \binits{B.}},
\bauthor{\bsnm{Wolf}, \binits{M.}},
\bauthor{\bsnm{Kumagai}, \binits{T.}}:
\batitle{Joule heating in single-molecule point contacts studied by tip-enhanced raman spectroscopy}.
\bjtitle{ACS Nano}
\bvolume{16}(\bissue{10}),
\bfpage{16443}--\blpage{16451}
(\byear{2022})
\end{barticle}
\endbibitem

\bibitem[\protect\citeauthoryear{Evers et~al.}{2020}]{EversRMP2020}
\begin{barticle}
\bauthor{\bsnm{Evers}, \binits{F.}},
\bauthor{\bsnm{Koryt\'ar}, \binits{R.}},
\bauthor{\bsnm{Tewari}, \binits{S.}},
\bauthor{\bsnm{Ruitenbeek}, \binits{J.M.}}:
\batitle{Advances and challenges in single-molecule electron transport}.
\bjtitle{Reviews of Modern Physics}
\bvolume{92},
\bfpage{035001}
(\byear{2020})
\end{barticle}
\endbibitem

\bibitem[\protect\citeauthoryear{Sergueev et~al.}{2005}]{SergueevPRL2005}
\begin{barticle}
\bauthor{\bsnm{Sergueev}, \binits{N.}},
\bauthor{\bsnm{Roubtsov}, \binits{D.}},
\bauthor{\bsnm{Guo}, \binits{H.}}:
\batitle{Ab initio analysis of electron-phonon coupling in molecular devices}.
\bjtitle{Physical Review Letters}
\bvolume{95},
\bfpage{146803}
(\byear{2005})
\end{barticle}
\endbibitem

\bibitem[\protect\citeauthoryear{Crampton et~al.}{2018}]{crampton2018junction}
\begin{barticle}
\bauthor{\bsnm{Crampton}, \binits{K.T.}},
\bauthor{\bsnm{Fast}, \binits{A.}},
\bauthor{\bsnm{Potma}, \binits{E.O.}},
\bauthor{\bsnm{Apkarian}, \binits{V.A.}}:
\batitle{Junction plasmon driven population inversion of molecular vibrations: a picosecond surface-enhanced raman spectroscopy study}.
\bjtitle{Nano letters}
\bvolume{18}(\bissue{9}),
\bfpage{5791}--\blpage{5796}
(\byear{2018})
\end{barticle}
\endbibitem

\bibitem[\protect\citeauthoryear{Gandra and Singamaneni}{2012}]{gandra2012bilayered}
\begin{barticle}
\bauthor{\bsnm{Gandra}, \binits{N.}},
\bauthor{\bsnm{Singamaneni}, \binits{S.}}:
\batitle{Bilayered raman-intense gold nanostructures with hidden tags (brights) for high-resolution bioimaging.}
\bjtitle{Advanced Materials (Deerfield Beach, Fla.)}
\bvolume{25}(\bissue{7}),
\bfpage{1022}--\blpage{1027}
(\byear{2012})
\end{barticle}
\endbibitem

\bibitem[\protect\citeauthoryear{Lin et~al.}{2018}]{lin2018electron}
\begin{barticle}
\bauthor{\bsnm{Lin}, \binits{L.}},
\bauthor{\bsnm{Zhang}, \binits{Q.}},
\bauthor{\bsnm{Li}, \binits{X.}},
\bauthor{\bsnm{Qiu}, \binits{M.}},
\bauthor{\bsnm{Jiang}, \binits{X.}},
\bauthor{\bsnm{Jin}, \binits{W.}},
\bauthor{\bsnm{Gu}, \binits{H.}},
\bauthor{\bsnm{Lei}, \binits{D.Y.}},
\bauthor{\bsnm{Ye}, \binits{J.}}:
\batitle{Electron transport across plasmonic molecular nanogaps interrogated with surface-enhanced raman scattering}.
\bjtitle{ACS Nano}
\bvolume{12}(\bissue{7}),
\bfpage{6492}--\blpage{6503}
(\byear{2018})
\end{barticle}
\endbibitem

\bibitem[\protect\citeauthoryear{Schmidt et~al.}{2017}]{schmidt2017linking}
\begin{barticle}
\bauthor{\bsnm{Schmidt}, \binits{M.K.}},
\bauthor{\bsnm{Esteban}, \binits{R.}},
\bauthor{\bsnm{Benz}, \binits{F.}},
\bauthor{\bsnm{Baumberg}, \binits{J.J.}},
\bauthor{\bsnm{Aizpurua}, \binits{J.}}:
\batitle{Linking classical and molecular optomechanics descriptions of sers}.
\bjtitle{Faraday Discussions}
\bvolume{205},
\bfpage{31}--\blpage{65}
(\byear{2017})
\end{barticle}
\endbibitem

\bibitem[\protect\citeauthoryear{Datta}{1997}]{datta1997electronic}
\begin{bbook}
\bauthor{\bsnm{Datta}, \binits{S.}}:
\bbtitle{Electronic Transport in Mesoscopic Systems}.
\bsertitle{Cambridge Studies in Semiconductor Physics and Microelectronic Engineering}.
\bpublisher{Cambridge University Press},
\blocation{Cambridge}
(\byear{1997})
\end{bbook}
\endbibitem

\bibitem[\protect\citeauthoryear{Toher and Sanvito}{2008}]{toher2008effects}
\begin{barticle}
\bauthor{\bsnm{Toher}, \binits{C.}},
\bauthor{\bsnm{Sanvito}, \binits{S.}}:
\batitle{Effects of self-interaction corrections on the transport properties of phenyl-based molecular junctions}.
\bjtitle{Physical Review B}
\bvolume{77}(\bissue{15}),
\bfpage{155402}
(\byear{2008})
\end{barticle}
\endbibitem

\bibitem[\protect\citeauthoryear{Gr{\"o}nbeck et~al.}{2000}]{gronbeck2000thiols}
\begin{barticle}
\bauthor{\bsnm{Gr{\"o}nbeck}, \binits{H.}},
\bauthor{\bsnm{Curioni}, \binits{A.}},
\bauthor{\bsnm{Andreoni}, \binits{W.}}:
\batitle{Thiols and disulfides on the au (111) surface: the headgroup- gold interaction}.
\bjtitle{Journal of the American Chemical Society}
\bvolume{122}(\bissue{16}),
\bfpage{3839}--\blpage{3842}
(\byear{2000})
\end{barticle}
\endbibitem

\bibitem[\protect\citeauthoryear{Tachibana et~al.}{2002}]{tachibana2002sulfur}
\begin{barticle}
\bauthor{\bsnm{Tachibana}, \binits{M.}},
\bauthor{\bsnm{Yoshizawa}, \binits{K.}},
\bauthor{\bsnm{Ogawa}, \binits{A.}},
\bauthor{\bsnm{Fujimoto}, \binits{H.}},
\bauthor{\bsnm{Hoffmann}, \binits{R.}}:
\batitle{Sulfur- gold orbital interactions which determine the structure of alkanethiolate/au (111) self-assembled monolayer systems}.
\bjtitle{The Journal of Physical Chemistry B}
\bvolume{106}(\bissue{49}),
\bfpage{12727}--\blpage{12736}
(\byear{2002})
\end{barticle}
\endbibitem

\bibitem[\protect\citeauthoryear{Nara et~al.}{2004}]{nara2004density}
\begin{barticle}
\bauthor{\bsnm{Nara}, \binits{J.}},
\bauthor{\bsnm{Higai}, \binits{S.}},
\bauthor{\bsnm{Morikawa}, \binits{Y.}},
\bauthor{\bsnm{Ohno}, \binits{T.}}:
\batitle{Density functional theory investigation of benzenethiol adsorption on au (111)}.
\bjtitle{The Journal of Chemical Physics}
\bvolume{120}(\bissue{14}),
\bfpage{6705}--\blpage{6711}
(\byear{2004})
\end{barticle}
\endbibitem

\bibitem[\protect\citeauthoryear{Hybertsen and Louie}{1986}]{hybertsen1986electron}
\begin{barticle}
\bauthor{\bsnm{Hybertsen}, \binits{M.S.}},
\bauthor{\bsnm{Louie}, \binits{S.G.}}:
\batitle{Electron correlation in semiconductors and insulators: Band gaps and quasiparticle energies}.
\bjtitle{Physical Review B}
\bvolume{34}(\bissue{8}),
\bfpage{5390}
(\byear{1986})
\end{barticle}
\endbibitem

\bibitem[\protect\citeauthoryear{Johnson and Christy}{1972}]{johnson1972optical}
\begin{barticle}
\bauthor{\bsnm{Johnson}, \binits{P.B.}},
\bauthor{\bsnm{Christy}, \binits{R.}}:
\batitle{Optical constants of the noble metals}.
\bjtitle{Physical Review B}
\bvolume{6}(\bissue{12}),
\bfpage{4370}
(\byear{1972})
\end{barticle}
\endbibitem

\bibitem[\protect\citeauthoryear{B\"urkle et~al.}{2012}]{BurklePRB2012}
\begin{barticle}
\bauthor{\bsnm{B\"urkle}, \binits{M.}},
\bauthor{\bsnm{Viljas}, \binits{J.K.}},
\bauthor{\bsnm{Vonlanthen}, \binits{D.}},
\bauthor{\bsnm{Mishchenko}, \binits{A.}},
\bauthor{\bsnm{Sch\"on}, \binits{G.}},
\bauthor{\bsnm{Mayor}, \binits{M.}},
\bauthor{\bsnm{Wandlowski}, \binits{T.}},
\bauthor{\bsnm{Pauly}, \binits{F.}}:
\batitle{Conduction mechanisms in biphenyl dithiol single-molecule junctions}.
\bjtitle{Physical Review B}
\bvolume{85},
\bfpage{075417}
(\byear{2012})
\end{barticle}
\endbibitem

\end{thebibliography}
